\documentclass[aps,prb,twocolumn,showpacs,superscriptaddress]{revtex4}
\usepackage{epsfig}
\usepackage{graphicx}
\usepackage{amsfonts}
\usepackage[figuresright]{rotating}
\usepackage{amssymb}
\usepackage{amsmath}
\usepackage{psfrag}
\usepackage{color}

\def\avg#1{\langle#1\rangle}
\def\Re{\rm{Re}}
\def\Im{\rm{Im}}
\def\be{\begin{equation}}
\def\ee{\end{equation}}
\def\bea{\begin{eqnarray}}
\def\eea{\end{eqnarray}}

\def\nn{\nonumber}

\def\Re{\mbox{Re}}

\begin{document}
\title{Unconventional states of bosons with synthetic spin-orbit coupling}
\author{ Xiangfa Zhou}
\affiliation{
Key Laboratory of Quantum Information, University of Science and Technology
of China, CAS, Hefei, Anhui 230026, China
}
\author{Yi Li}
\affiliation{Department of Physics, University of California, San Diego,
CA 92093, USA}
\author{Zi Cai}
\affiliation{Physics Department, Arnold Sommerfeld Center for
Theoretical Physics, and Center for NanoScience,
Ludwig-Maximilians-Universit\"{a}t M\"{u}nchen, D-80333 M\"{u}nchen,
Germany}
\author{Congjun Wu}
\affiliation{Department of Physics, University of California, San Diego,
CA 92093, USA}

\begin{abstract}
Spin-orbit coupling with bosons gives rise to novel properties that are absent
in usual bosonic systems.
Under very general conditions, the conventional ground state wavefunctions
of bosons are constrained by the ``no-node'' theorem to be positive-definite.
In contrast, the linear-dependence of spin-orbit coupling leads to
complex-valued condensate wavefunctions beyond this theorem.
In this article, we review the study of this class of unconventional
Bose-Einstein condensations focusing on their topological properties.
Both the 2D Rashba and 3D $\vec \sigma \cdot \vec p$-type Weyl spin-orbit
couplings give rise to Landau-level-like quantization of single-particle
levels in the harmonic trap.
The interacting condensates develop the half-quantum vortex structure
spontaneously breaking time-reversal symmetry and exhibit topological spin
textures of the skyrmion type.
In particular, the 3D Weyl coupling generates topological defects in the
quaternionic phase space as an SU(2) generalization of the usual U(1) vortices.
Rotating spin-orbit coupled condensates exhibit rich vortex structures
due to the interplay between vorticity and spin texture.
In the Mott-insulating states in optical lattices, quantum
magnetism is characterized by the Dzyaloshinskii-Moriya type exchange
interactions.
\end{abstract}

\pacs{Keywords: Bose-Einstein condensation, spin-orbit coupling,
Landau level, time-reversal symmetry, spin texture, skyrmion,
quaternion, Dzyaloshinskii-Moriya.
}
\maketitle

\section{Introduction}
\label{sect:intro}
Spin-orbit (SO) coupling plays an important role in interdisciplinary
areas of physics.
In quantum mechanics, SO coupling arises from the relativistic effect
as a low energy approximation to the Dirac equation.
Its semi-classic picture is the Thomas precession that electron spin moment
couples to a velocity-dependent effective magnetic field generated by
the Lorentz transformation of the electric field.
In atomic physics, SO coupling constitutes one of the basic elements
to the formation of the atomic structures.
The development in condensed matter physics shows that SO coupling
is indispensable for important phenomena ranging from spintronics
\cite{ifmmodecheckZelsevZfiutiifmmodeacutecelsecfi2004},
anomalous Hall effects \cite{Nagaosa2010,Xiao2010}, spin Hall effects
\cite{Dyakonov1971,Hirsch1999,Murakami2003,Sinova2004},
to topological insulators \cite{Hasan2010,Qi2011}.
In particular, topological insulators have become a major
research focus of current condensed matter physics.

Most current studies of SO coupling are considered for fermionic systems
of electrons.
On the other hand, the ultra-cold atomic systems have opened up a whole
new opportunity to explore novel states of matter that are not easily
accessible in usual condensed matter systems.
In particular, it currently becomes experimentally possible to implement
various kinds of SO coupled Hamiltonians in ultracold atomic gases for
both fermions and bosons \cite{Galiski2013,Lin2009,Lin2009a,Lin2011,Zhang2012a,
Wang2012,cheuk2012,Qu2013}.
The high controllability of these systems makes them an ideal platform to
explore novel SO coupled physics with bosons.

An important property of bosons is the ``no-node" theorem, which
states that in the coordinate representation the many-body ground state
wavefunctions are positive-definite \cite{Feynman1972}.
This theorem is valid under very general conditions such as the Laplacian
type kinetic energy, arbitrary single-particle potentials, and
coordinate-dependent two-body interactions.
It applies to most of the known ground states of bosons, including
Bose-Einstein condensations (BECs), Mott insulators, and supersolids.
Technically, it indicates that the ground state wavefunctions of bosons can
be reduced to positive-definite distributions, and thus imposes a strong
constraint on bosonic states.
For example, it rules out the possibility of time-reversal (TR) symmetry
breaking ground states in traditional boson systems.
Considerable efforts have been made in exploring unconventional BECs beyond
the ``no-node'' theorem \cite{Wu2009}.
One way is the meta-stable state of bosons in high orbital
bands in optical lattices \cite{Isacsson2005,Liu2006,Kuklov2006,Cai2011},
because the ``no-node'' theorem does not apply to excited states.
Unconventional BECs with complex-valued condensate wavefunctions have been
experimentally realized \cite{Muller2007,Wirth2010,Olschlager2011}.

The kinetic energies of SO coupled systems are no longer Laplacian but
linearly depend on momentum, which invalidates the necessary conditions
for the ``no-node'' theorem as pointed out in Ref. [\onlinecite{Wu2009}].
This provides another way towards unconventional BECs.
For instance, the Rashba SO coupled BECs were early investigated
for both isotropic \cite{Wu2011} and anisotropic cases \cite{Stanescu2008}.
In the isotropic case, Rashba coupling leads to degenerate single-particle
ground states along a ring in momentum space whose radius $k_{so}$
is proportional to the SO coupling strength.
Such a momentum scale is absent in usual BECs in which bosons are condensed
to the zero momentum state, and thus bears certain similarities to Fermi
momentum in fermion systems.
If interaction is spin-independent, the condensates are frustrated in the
free space at the Hartree-Fock level, and quantum zero point energy selects
a spin-spiral state based on the ``order-from-disorder'' mechanism.
Imposing the trapping potential further quantizes the motion around
the SO ring which leads to Landau level type quantization of the
single-particle levels.
Under interactions, condensates spontaneously break TR symmetry
exhibiting topologically non-trivial spin textures of the skyrmion type
\cite{Wu2011}.
All these features are beyond the framework of ``no-node'' theorem.

Recently, SO coupled systems with ultra-cold bosons have aroused a great
deal of research interest both experimental and theoretical \cite{Galiski2013}.
Experimentally, pioneered by Spielman's group \cite{Lin2009,Lin2009a,
Lin2011}, BECs with SO coupling in the anisotropic 1D limit have been
realized by engineering atom-laser interactions through Raman processes
\cite{Lin2009,Lin2009a,Lin2011,Zhang2012a,Wang2012,Qu2013}.
Condensations at finite momenta and exotic spin dynamics have been observed
\cite{Lin2009,Lin2009a,Lin2011,Zhang2012a,Wang2012,cheuk2012,Qu2013}.
Various experimental schemes have proposed to realize
the isotropic Rashba SO coupling \cite{Jaksch2003,Ruseckas2005,Osterloh2005,
Campbell2011,Juzeliunas2011,Dalibard2011}.
On the side of theory, the Rashba SO coupled bosons have been extensively
investigated under various conditions, including the exotic spin structures
in the free space, spin textures in harmonic traps,
vortex structures in rotating traps, and the SO coupled quantum magnetism
in the Mott-insulating states
\cite{Stanescu2008,Wu2011,Wang2010,Ho2011,Anderson2011,
Burrello2010,Burrello2011,Kawakami2011,Li2011,Sinha2011,Xu2011a,Yip2011,Zhu2011,
Anderson2012,Anderson2012a,Barnett2012,Deng2012,Grass2012,He2012b,Hu2012a,
LiYun2012,Ramachandhran2012,Ruokokoski2012,Sedrakyan2012,Xu2012,
Xu2012b,Zhang2012,Zhang2012c,Zhang2012d,Zheng2012,Cui2012}.
Furthermore, a recent progress shows that a 3D  $\vec\sigma\cdot \vec p$-type
SO coupling can also been implemented with atom-laser interactions
\cite{Li2012c,Anderson2012a}.
This is a natural symmetric extension of Rashba SO coupling to 3D dubbed
Weyl SO coupling below due to its similarity to the
relativistic Hamiltonian of Weyl fermions \cite{Weyl1929}.
The Weyl SO coupled BECs have also been theoretically investigated
\cite{Li2012b,Kawakami2012,Zhang2013,Anderson2012a}.

In addition to the ultra-cold atom systems, recent progress in condensed
matter systems has also provided an SO coupled boson system of excitons.
Excitons are composite objects of conduction electrons and valence holes,
both of their effective masses are small, thus relativistic SO coupling
exists in the their center of mass motion.
The effects of SO coupling on exciton condensations have
been theoretically investigated \cite{Yao2008,Wu2011}, including
the spin texture formations \cite{Wu2011}.
An important experiment progress has been achieved in Butov's group
\cite{High2011,High2012}, that spin textures in a cold
exciton gas have been observed in GaAs/AlGaAs coupled quantum
wells from the photoluminescence measurement.

In the rest of this article, we review the current theoretical progress
of studying SO coupled bosons including both the 2D Rashba and 3D Weyl
SO couplings.
Our emphases will be on the non-trivial topological properties which
are absent in conventional BECs.
The single-particle spectra will be reviewed in Section \ref{sect:single}.
They exhibit a similar structure to the Landau level
quantization in the sense that the dispersion with angular momentum is
strongly suppressed by SO couplings \cite{Wu2011,Burrello2010,Burrello2011,Li2012c,Hu2012a,
Sinha2011,Ghosh2011,Li2012b}.
However, a crucial difference from the usual magnetic Landau levels
is that these SO coupling induced Landau levels maintain time-reversal
symmetry, and thus their topology belongs to the $Z_2$ class
\cite{Wu2011}.
The interplay between interactions and topology gives rise to a variety of
topological non-trivial condensate configurations and spin textures,
which will be reviewed in Section \ref{sect:texture}.
In particular, the 3D condensates with the Weyl coupling exhibit
topological defects in the quaternionic phase space.
It is exciting to find an application of the beautiful mathematical
concept of quaternions.
In Section \ref{sect:vort}, we review the SO coupled BECs in rotating
traps, which are subject to both the Abelian
vector potential due to the Coriolis force and the non-Abelian one
from SO coupling.
The combined effects of the vorticity and spin topology lead to rich
structures\cite{ZhouXF2011,Radic2011,Xu2011,Liu2012,Zhao2013},
including half quantum vortex lattices, multi-domain of plane-wave states,
and giant vortices.
In Section \ref{sect:Strong}, we summarize the current progress on strongly
correlated spin-orbit coupled systems \cite{Ramachandhran2013,grass2012b,grass2012c,palmer2011,komineas2012}.
Furthermore, in the strongly correlated Mott-insulators, SO coupling
effects exhibit in the quantum magnetism as
the Dzyaloshinskii-Moriya type exchange interactions
\cite{Cai2012,Cole2012,Radic2012,Gong2012,Mandal2012,grass2011}, which
will be reviewed in Section \ref{sect:Mott}.
Conclusions and outlooks are presented in Section \ref{sect:conclusion}.

Due to the rapid increasing literatures and the limit of space, we will not
cover other important topics, such as SO coupled fermions \cite{Ghosh2011,
Iskin2011,Jiang2011,Li2011,Zhou2011,Doko2012,He2012,He2012a,Hu2012,
Liu2012a,Maldonado-Mundo2012,Martone2012,Orth2012,Vyasanakere2012,
Zhang2012b,cheuk2012}, the SO coupled dipolar bosons \cite{Deng2012,Zhao2013}.
and the proposals for experimental implementations
\cite{Jaksch2003,Ruseckas2005,Osterloh2005,
Campbell2011,Juzeliunas2011,Dalibard2011,tagliacozzo2012,goldman2013}.

\section{The SO coupled single-particle spectra and
the Landau level quantization}
\label{sect:single}

We begin with the single-particle properties.
Consider the following Hamiltonian of 2D two-component atomic gases
with an artificial Rashba SO coupling defined as
\bea
H_0^{2D,R}&=& \frac{\vec{p}^2 }{2M} + V_{tp}(\vec{r}) - \lambda_R
(\sigma_x p_y -\sigma_y p_x),
\label{eq:single_rashba}
\eea
where $\vec p=-i\hbar \vec{\nabla}$; 
the pseudospin components $\uparrow$ and $\downarrow$ refer to two
different internal atomic components; $\lambda_R$ is the Rashba SO coupling
strength with the unit of velocity;
$V_{tp}(\vec{r})=\frac{1}{2}M \omega^2 r^2$ is the external
trapping potential, and $\omega$ is the characteristic frequency of the trap.
Another SO coupled Hamiltonian will be considered is
the 3D Weyl SO coupling defined as
\bea
H_{0}^{3D,W}&=&
\frac{\vec{p}^2 }{2M} + V_{tp}(\vec{r}) - \lambda_{W}
\vec{\sigma} \cdot \vec{p},
\label{eq:single_3D}
\eea
where $\lambda_W$ is the SO coupling strength.

Even though we will mostly consider bosons for the Hamiltonians of
Eq. (\ref{eq:single_rashba}) and Eq. (\ref{eq:single_3D}), they possess
a Kramer-type TR symmetry $T=i\sigma_y C$ satisfying
$T^2=-1$, where $C$ is the complex conjugate operation.
At the single particle level, there is no difference between bosons
and fermions.
Both Hamiltonians are rotationally invariant but break the inversion symmetry.
The 2D Hamiltonian Eq. (\ref{eq:single_rashba}) still possesses
the reflection symmetry with respect to any vertical plane passing the
center of the trap.
For the 3D Hamiltonian Eq. (\ref{eq:single_3D}), no reflection symmetry
exists.

These two typical types of SO interactions have received a lot of
attention recently in the community of ultra-cold atoms due to their close
connection to condense matter physics.
There have already been great experimental efforts on
realizing spin-orbit coupling through atom-light
interactions \cite{Lin2009,Lin2009a,Lin2011,Wang2012,Zhang2012,Qu2013}.
In fact, several proposals for experimental implementations
of Eq. (\ref{eq:single_rashba}) and Eq. (\ref{eq:single_3D})
have appeared in literatures \cite{Ruseckas2005,Juzeliunas2011,
Anderson2012a,Dalibard2011,LiYun2012}.

In this section, we review the single-particle properties of Eqs.
(\ref{eq:single_rashba}) and (\ref{eq:single_3D}) focusing on their
topological properties.
In Sec. \ref{sect:berry},  their Berry phase structures in
momentum space are presented.
When the quadratic harmonic trap potential is imposed, the Landau-level
type quantization on the energy spectra appears with TR symmetry as
shown in Sec. \ref{sect:landau}.
This Landau level quantization provides a
clear way to understand novel phases of bosons after turning
on interactions.
In Sec. \ref{sect:wf}, wavefunctions of the lowest Landau levels
of Eq. (\ref{eq:single_rashba}) and Eq. (\ref{eq:single_3D}) are
explicitly provided.
The topology of these Landau level states are
reviewed through edge and surface spectra in
Sec. \ref{sect:topo}.

\subsection{Berry connections in momentum space}
\label{sect:berry}

Both Eq. (\ref{eq:single_rashba}) and Eq. (\ref{eq:single_3D}) possess
non-trivial topology in momentum space.
Let us begin with the 2D Rashba Hamiltonian Eq.
(\ref{eq:single_rashba})
in the free space, i.e., $V_{tp}=0$.
Its lowest single-particle states in free space is not located at the
origin of momentum space but around a ring with the radius
$k^R_{so}=M\lambda_R/\hbar$.
The spectra read
\bea
\epsilon_{\pm} (\vec k)= \frac{\hbar^2}{2 M} (k \mp
k^R_{so})^2,
\label{eq:rashba_spectrum}
\eea
where $\pm$ refer to the helicity eigenvalues of the operator
$\vec \sigma \cdot (\hat{k} \times \hat z)$.
The corresponding two-component spin wavefunctions of plane-wave
states $|\psi_{\vec k\pm}\rangle$
are solved as
\bea
|\psi_{\vec k\pm} \rangle = \frac{1}{\sqrt 2}
\left(\begin{array}{c}
e^{-i \frac{\phi_{\vec{k}}}{2}}\\
\mp i e^{i \frac{\phi_{\vec{k}}}{2}}
\end{array}
\right),
\eea
where $\phi_{\vec{k}}$ is the azimuthal angle of $\vec k$ in the $xy$-plane.

For bosons, the lower energy branch states with a fixed helicity are important.
The Berry connection $\vec A(\vec k)$ of  positive helicity states
$\psi_+(\vec k)$ is defined as
\bea
\vec A(\vec k)= \avg{\psi_{{\vec{k}}+}|i \vec \nabla_k |\psi_{{\vec{k}}+}}
=\frac{1}{2k} \hat e_{\phi_{\vec{k}}},
\label{eq:A_rashba}
\eea
where $\hat e_{\phi_{\vec{k}}}$ is the unit vector along the azimuthal direction.
The Berry curvature $F_{ij}$ is defined as
$F_{ij}(\vec k)=\partial_{k_i} A_j (\vec k) -\partial_{k_j} A_i (\vec k)$.
For a loop winding around the origin $\vec k=(0,0)$, the Berry phase is
\bea
\oint d \vec k \cdot \vec A(\vec k)= \pi.
\eea
This is because a two-component spinor after rotating 360$^\circ$
does not return to itself but acquires a minus sign.
Consequently, $F_{ij}(\vec k)$ is zero everywhere except
contributing a $\pi$-flux at the origin of momentum space.

Next we consider the 3D generalization of the Rashba SO coupling
of the $\vec \sigma \cdot \vec p$ type, i.e., the Weyl coupling.
Now in the free space without the trap, the lowest energy single-particle
states are located around a sphere in momentum space with the
radius also denoted as $k^W_{so}$ with
the value of $k^W_{so}=M\lambda_W/\hbar$, and the spectra are
\bea
\epsilon_{\pm} (\vec k)= \frac{\hbar^2}{2 M} \Big(k \mp k^W_{so}\Big)^2,
\label{eq:3dSO_spectrum}
\eea
where the subscripts $\pm$ refer to the helicity eigenvalues of the operator
$\vec \sigma \cdot \hat{k}$.
The corresponding eigenstates are solved as
\bea
| \psi_{\vec k -} \rangle =
\left (  \begin{array}{c}
-\sin \frac{\theta_{\vec{k}}}{2} \\
\cos \frac{\theta_{\vec{k}}}{2} e^{i \phi_{\vec{k}}}
\end{array}
\right ), \ \ \,
| \psi_{\vec k +} \rangle =
\left (  \begin{array}{c}
\cos \frac{\theta_{\vec{k}}}{2} \\
\sin \frac{\theta_{\vec{k}}}{2} e^{i \phi_{\vec{k}}}
\end{array}
\right ),
\eea
where $\phi_{\vec{k}}$ and $\theta_{\vec{k}}$ are the azimuthal and polar angles of $\vec k$
in the spherical coordinates.
The Berry connection of the positive helicity states $\psi_{\vec k,+}$ is
\bea
\vec A(\vec k)= \frac{1}{2} \tan \frac{\theta_{\vec{k}}}{2} \hat e_{\phi_{\vec{k}}},
\label{eq:monopole}
\eea
which is the vector potential for a unit magnetic monopole
located at the origin of momentum space, and $\hat e_{\phi_{\vec{k}}}$
is the azimuthal direction of $\vec k$.
Defining $B_i=\frac{1}{2}\epsilon_{ijl} F_{jl}$, the
corresponding Berry curvature is
$\vec B (\vec k) =  \frac{1}{2k^2} \hat e_{\vec{k}}$,
where $\hat e_{\vec{k}}$ is the radial direction of $\vec k$.

\subsection{Landau-level quantization in the harmonic trap from
SO couplings}
\label{sect:landau}

The SO couplings in Eqs. (\ref{eq:single_rashba}) and
(\ref{eq:single_3D})
introduce a SO length scale even in the free space defined as $l_{so}=1/k_{so}$.
Here and the following, we omit the superscripts of $k^R_{so}$ and $k^W_{so}$
without loss of generality. 
The physical meaning of $l_{so}$ is as follows: the low energy states
of Eqs.  (\ref{eq:single_rashba}) and (\ref{eq:single_3D}) are not of
long-wave length as usual but featured with large magnitude of momentum
depending on the SO coupling strength.
$l_{so}$ is the length scale of wavepackets that can be formed by using
the low energy states on the SO ring of Eq. (\ref{eq:single_rashba}) or
the SO sphere of Eq. (\ref{eq:single_3D}).
On the other hand, in the typical experimental setup with ultra-cold quantum
gases, a harmonic trap is used to confine atoms.
The trapping potential $V(\vec{r})=\frac{1}{2}m \omega^2r^2$ introduces
another length scale as $l_T=\sqrt{\hbar / (m\omega)}$
as the typical system size.
The trap energy scale is $E_{tp}=\hbar \omega$.

It is useful to define a dimensionless parameter $\alpha=l_T/l_{so}$
to describe the relative strength of SO coupling with respect to
the trapping potential.
Physically, $\alpha$ is the number of wavepackets which can be packed
in the trap length.
In the limit of large values of $\alpha$, the trapping potential gives rise
to Landau level type quantizations in both 2D and 3D spin-orbit coupling
systems \cite{Wu2011,Hu2012,Ghosh2011,LiYun2012}.

The terminology of Landau levels in this section is generalized from
the usual 2D magnetic case as: {\it topological} single-particle
level structures labeled by {\it angular momentum} quantum numbers with
flat or nearly flat spectra.
On open boundaries, Landau levels systems develop gapless surface
or edge modes which are robust against disorders.
We will see that the low energy states of
both Eq. (\ref{eq:single_rashba}) and Eq. (\ref{eq:single_3D})
satisfy this criterion in the case $\alpha\gg 1$.

\subsubsection{TR invariant Landau levels from the 2D Rashba SO coupling}
\label{sect:2DLL}
Let us briefly recall the usual 2D Landau level arising from the magnetic
field.
In the symmetric gauge with $\vec A=\frac{1}{2} B \hat z \times \vec r$,
its Hamiltonian is simply equivalent to a 2D harmonic oscillator
in a rotating frame as
\bea
H_{2D,LL}=\frac{(\vec p-\frac{e}{c}\vec A)^2}{2M}
=\frac{p^2}{2M}+\frac{1}{2}M \omega^2r^2 -\omega L_z,
\label{eq:2D_LL}
\eea
where $L_z=xp_y-yp_x$ and $\omega=\frac{|eB|}{Mc}$.
Inside each Landau level, the spectra are degenerate with respect to
the magnetic quantum number $m$.
Non-trivial topology of Landau levels comes from the fact
that $m$ does not take all the integer values.
For example, in the lowest Landau level, $m$ starts from $0$ and runs all the
positive integer number.
This chiral feature is a TR symmetry breaking effect due to
the magnetic field.

Next let us consider the 2D Rashba SO coupling of Eq.
(\ref{eq:single_rashba})
in the limit of $\alpha\gg 1$.
The physics is most clearly illustrated in momentum representation.
After projected into the low energy sector of the positive helicity
states, the harmonic potential in momentum space becomes a Laplacian
subjected to the Berry connection as
\bea
V_{tp}(\vec{\nabla}_{\vec{k}})=\frac{M}{2} \omega^2 (i\vec{\nabla}_{\vec k} - \vec A_{\vec k})^2,
\eea
where $\vec A$ is given in Eq. (\ref{eq:A_rashba}) with a $\pi$-flux
at the origin of the 2D $k_x$-$k_y$ plane.
In momentum space, the trapping potential quantizes the motion on the low energy
spin-orbit ring with radius $k_{so}$, and is mapped to a planar rotor
problem.
The moment of inertial in momentum space is $I_k=k_{so}^2M_k$
where $M_k$ is the mass in momentum space defined as $M_k=1/(M\omega^2)$,
and the angular dispersion of energy is
$E_{agl}(j_z)=\hbar^2 j_z^2 /2I_k=\frac{1}{2\alpha^2} j_z^2 E_{tp}$.
Due to the $\pi$-flux phase at $\vec k=(0,0)$, $
j_z$ is quantized to half-integer values.
On the other hand, the radial component of the trapping potential
in momentum space is just the kinetic energy for the positive
helicity states
\bea
H_K=\frac{1}{2}M_k \omega^2 (k-k_{so})^2.
\eea
For states near the low energy spin-orbit ring,
the radial motion can be approximated as 1D harmonic oscillations,
and the energy gap remains as $\hbar\omega$.
Combining the radial and angular dispersions together, we arrive at
\bea
E_{n_r,j_z}\approx \Big\{n_r+ \frac{j_z^2}{2\alpha^2}
+\frac{1}{2}(1- \alpha^2) \Big \}E_{tp},
\label{eq:2D_Rashba_gap}
\eea
where $n_r$ is the radial quantum number, and $\frac{1}{2}(1-\alpha^2)$
is the constant of zero point energy.

The degeneracy over angular momentum quantum numbers is a main feature of
Landau level quantization.
For the Hamiltonian Eq. (\ref{eq:single_rashba}), although its spectra
Eq. (\ref{eq:2D_Rashba_gap}) are not exactly flat with respect to $j_z$,
they are strongly suppressed at $\alpha\gg 1$, thus
these low energy levels are viewed as Landau levels.
The radial quantum number $n_r$ serves as the Landau level index
and the gaps between Landau levels are roughly $E_{tp}$.
For states in the $n_r$-th Landau level with $|j_z|\le
\sqrt 2 \alpha$, their energies remain lower than the bottom of
the next Landau level, thus they can be viewed as gapped bulk states.
Actually, the similarities of these SO coupled states to Landau
levels are more than just spectra flatness but their
non-trivial topology  will be explained in Sec. \ref{sect:topo}.

\subsubsection{3D Landau levels from the Weyl SO coupling}
The 3D Landau level systems are not as well-known as the 2D case
of Eq. (\ref{eq:2D_LL}).
Recently, a large progress has been made in generalizing Eq.
(\ref{eq:2D_LL}) to 3D with exactly flat energy dispersions
\cite{Li2011,Li2012d}.
In particular, they can be constructed with the full 3D rotation
symmetry by coupling spin-$\frac{1}{2}$ fermions with the SU(2) gauge
potential.
The Hamiltonian is equivalent to a 3D harmonic oscillator plus SO
coupling as
\bea
H_{3D,LL}=\frac{p^2}{2M}
+\frac{1}{2}M\omega^2 r^2 -\omega \vec L \cdot \vec \sigma.
\label{eq:3D_LL}
\eea
Excitingly, the lowest Landau level wavefunctions of Eq.
(\ref{eq:3D_LL})
possess elegant analytic properties, satisfying the Cauchy-Riemann-Fueter
condition of quaternionic analyticity.
Just like that the complex analyticity is essential for the construction of
fractional quantum Hall Laughlin states, the quaternionic analyticity
is expected to play an important role in high dimensional
fractional topological states.
These 3D Landau level states preserve both TR and parity symmetry.
The 3D Landau levels have also been generalized to the relativistic
Dirac particles \cite{Li2012}.

Now let us come back to the Hamiltonian of Eq. (\ref{eq:single_3D}) with the
3D Weyl SO coupling and a trap potential.
The parallel analysis to the 2D Rashba case applies.
Again, in the limit of $\alpha\gg 1$, after the projection into the sector
of the positive helicity states, the trap potential becomes
$V_{tp}(\vec{\nabla}_{\vec{k}})
=\frac{1}{2}M (i \vec \nabla_{\vec{k}} -\vec A_{\vec{k}})^2$ and $\vec A_{\vec{k}}$ takes the
form of a magnetic monopole one in Eq. (\ref{eq:monopole}).
The problem is reduced to a spherical rotor problem in momentum space
on the low energy SO sphere with the radius $k_{so}$.
The monopole structure of the Berry connection quantizes the total angular
momentum $j$ to half-integer values.
Similarly to the 2D Rashba case, the low energy spectra are approximated as
\bea
E_{n_r,j,j_z}\approx \Big\{ n_r+\frac{j(j+1)}{2\alpha^2}
+\frac{1}{2}(1-\alpha^2) \Big\} E_{tp}
\label{eq:3D_SO_gap}.
\eea
Again the angular dispersion is strongly suppressed by SO coupling
at $\alpha\gg 1$.
These spectra exhibit quasi-degeneracy over the 3D angular momentum
good quantum numbers of $j$ and $j_z$, and thus can be viewed as a
3D Landau level quantization with TR symmetry.
The length scale of these Landau level states is also the SO
length $l_{so}$.
Topological properties of these Landau level states will be
studied in Sec. \ref{sect:topo}.

\subsection{Lowest Landau level wavefunctions and parent Hamiltonians}
\label{sect:wf}

The Landau level energy spectra of Eq. (\ref{eq:2D_Rashba_gap}) in 2D and Eq.
(\ref{eq:3D_SO_gap}) in 3D are not exactly flat but with weak
dispersions over angular momentum quantum numbers.
Nevertheless, parent Hamiltonians based on slight modification
on Eqs. (\ref{eq:single_rashba}) and (\ref{eq:single_3D}) can be constructed.
Their lowest Landau level spectra are exactly flat and their wavefunctions
can be solved analytically as shown in Eqs. (\ref{eq:2D_LL_rashba})
and (\ref{eq:3D_WF}) below.
These wavefunctions maintain TR symmetry but break parity.
In the limit of $\alpha\gg 1$ and for Landau level states with
angular momenta $|j_z|<\alpha$ in 2D or $j<\alpha$ in 3D,
the lowest Landau level wavefunctions
of Eqs. (\ref{eq:single_rashba}) and (\ref{eq:single_3D}) are
well approximated by these expressions.

For the 2D case, the parent Hamiltonian is just
\bea
H_{0}^{2D,P}&=& H^{2D,R}_0 - \omega L_z \sigma_z,
\label{eq:2D_reduce}
\eea
where $L_z=xp_y-yp_x$ and the coefficient $\omega$ is the same as
the trap frequency.
As shown in Ref. [\onlinecite{Li2012c}], its lowest Landau level wavefunctions
are solved as
\bea
\psi_{2D,j_z}^{LLL}(r,\phi)=
e^{-\frac{r^2}{2l_T^2}}
\left( \begin{array}{c}
e^{im\phi} J_m(k_{so} r) \\
-e^{i(m+1)\phi} J_{m+1} (k_{so} r)
\end{array} \right),
\label{eq:2D_LL_rashba}
\eea
where $\phi$ is the azimuthal angle; $j_z=m+\frac{1}{2}$;
$J_m$ is the $m$-th order Bessel function.
The lowest Landau level energy is exactly flat as
$E^{LLL}=(1-\frac{\alpha^2}{2})\hbar \omega$.

In the case of $\alpha\gg 1$ and for small values of $|j_z|\approx m<\alpha$,
the decay of the wavefunctions Eq. (\ref{eq:2D_LL_rashba}) is controlled
by the Bessel functions rather than the Gaussian factor.
Their classic orbit radiuses scale as $\rho_{c,j_z}\approx m l_{so}$.
Since $L_z$ linearly depends on $\rho_{c,j_z}$, the effect of the $L_z\sigma_z$
term compared to that of the Rashba one is at the order of $\rho_{c,j_z}
\omega/\lambda_R\approx m/\alpha^2 \ll 1$.
Thus Eq. (\ref{eq:2D_reduce}) is simply reduced to Eq.
(\ref{eq:single_rashba})
whose lowest Landau level wavefunctions are well approximated by
Eq. (\ref{eq:2D_LL_rashba}).
In this case, the length scale of Landau level states is determined by
the SO length $l_{so}$ instead of the trap length $l_T$.
The reason is that these Landau levels are composed from plane-wave
states with a fixed helicity on the low energy Rashba ring.
The confining trap further opens the gap at the order of $\hbar\omega$
between SO coupled Landau levels.

On the contrary, in the opposite limit, i.e., $|j_z|\approx m \gg \alpha^2$,
the $L_z\sigma_z$ term dominates and the Rashba term can be neglected.
In this case, Eq. (\ref{eq:2D_reduce}) is reduced into
$p^2/2M +\frac{1}{2}M\omega^2 r^2-\omega L_z \sigma_z$
with $\sigma_z$ conserved.
In each spin eigen-sector, it is just the usual Landau level Hamiltonian
in the symmetric gauge with opposite chiralities for spin up and down,
respectively.
Nevertheless, at $m\gg\alpha^2$, the approximation of the projection
into the Rashba ring for Eq. (\ref{eq:single_rashba}) is not valid,
and the eigenstates are no longer Landau levels.
For the intermediate values of $\alpha<|j_z|<\alpha^2$, the physics
is a crossover between the above two limits.

Following the same logic, the 3D parent Hamiltonian with exactly
flat SO coupled Landau levels is
\bea
H_{0}^{3D,P}=H^{3D,W}_0-\omega \vec L \cdot \vec \sigma,
\label{eq:3D_reduce}
\eea
where $\vec L=\vec r\times \vec p$ is the 3D orbital angular momentum,
and the coefficient of the $\vec L\cdot \vec \sigma$ term is the
same as the trap frequency.
Again, as shown in Ref. [\onlinecite{Li2012c}], the lowest Landau level
wavefunctions of Eq. (\ref{eq:3D_reduce}) are solved analytically as
\bea
\psi^{LLL}_{3D, jj_z}(\vec r)&=& e^{-\frac{r^2}{2l_T^2}} \Big\{ j_l(k_{so} r)
Y_{+,j,l,j_z} (\Omega_r)
+i j_{l+1}(k_{so} r) \nn \\
&\times& Y_{-,j, l+1, j_z} (\Omega_r) \Big\},
\label{eq:3D_WF}
\eea
where $j_l$ is the $l$-th order spherical Bessel function;
$Y_{\pm,j, l,j_z}$'s are the SO coupled spherical harmonics
with total angular momentum quantum numbers $j=l\pm\frac{1}{2}$
and $j_z$, which are composed of the spherical harmonics
$Y_{lm}$ and spin-$\frac{1}{2}$ spinors.
These lowest Landau level states are degenerate over all the
values of $(jj_z)$ with $E^{LLL}=(\frac{3}{2}-\frac{\alpha^2}{2})\hbar \omega$.

Following the same reasoning as in the 2D case,
in the limit of $\alpha\gg 1$, we can divide the lowest Landau level states
of Eq. (\ref{eq:3D_WF}) into three regimes as $j<\alpha$, $j\gg\alpha^2$,
and $\alpha<j<\alpha^2$, respectively.
At $j<\alpha$, the classic orbit radius scales as $r_{c,j}/l_T \approx
\frac{j}{\alpha}$, and thus $\vec \sigma \cdot \vec L$
comparing with $\vec \sigma \cdot \vec p$
is a perturbation at the order of $j/\alpha^2\ll 1$.
In this regime, the lowest Landau level wavefunctions of Eq. \ref{eq:single_3D}
are well approximated by Eq. (\ref{eq:3D_WF}).
On the contrary, in the regime of $j\gg \alpha^2$, $\vec \sigma \cdot \vec L$
dominates over $\vec \sigma\cdot \vec p$, thus the
eigenstates of Eqs. (\ref{eq:3D_reduce}) and (\ref{eq:single_3D})
are qualitatively different.
In this case, Eq. (\ref{eq:3D_reduce}) is reduced to the 3D
Landau level Hamiltonian Eq. (\ref{eq:3D_LL}).

\subsection{The $Z_2$-stability of helical edge and surface states}
\label{sect:topo}

Non-trivial topology of the 2D Landau level manifests from the appearance
of robust gapless edge states.
The classic radius $r_c$ of each Landau level state expands as $m$ increases.
For example, in the lowest Landau level, $r_c=\sqrt m l_B$ where
$l_B=\sqrt{\frac{\hbar c}{|eB|}}$ is the cyclotron radius.
With an open boundary, as $m$ becomes large enough, states are pushed to
the boundary \cite{halperin1982}.
Unlike the flat bulk spectra, the edge spectra are dispersive,
always increasing with $m$, and thus are chiral and robust against
external perturbations.
Each Landau level contributes one branch of chiral edge modes.
If the system is filled with fermions, when chemical potential $\mu$ lies
in the gap between Landau levels, the chiral edge states give rise
to the quantized charge transport.

For Landau levels of the Rashba SO coupling in Eq. (\ref{eq:single_rashba}),
a marked difference is that these states are TR invariant.
The angular momentum $j_z$ in Eq. (\ref{eq:2D_Rashba_gap})
takes all the half-integer values as $j_z=\pm\frac{1}{2},\pm\frac{3}{2},
..., \pm (m+\frac{1}{2}), ...$, and thus these states are helical
instead of chiral.
Since the system described by Eq. (\ref{eq:single_rashba}) does not
possess translation symmetry, the usual method of calculating topological
index based on lattice Bloch wave structures in Brillouin zones
does not work \cite{kane2005,fu2007,moore2007,roy2010}.

Nevertheless, the non-trivial topology should exhibit on the robustness of
edge states.
The trap length $l_T$ can be used as the sample size by imposing an open
boundary condition at $r=l_T$.
States with $|j_z|< \alpha$ are bulk states localized within the region
of $r< l_T$.
States with $|j_z|\sim \alpha$ are pushed to the boundary, whose spectra
disperse to high energy rapidly.
For a given energy $E$ lying between Landau level gaps,
each Landau level with bulk energy blow $E$ contributes to a pair of degenerate
edge modes $\psi_{\pm j_z}$ due to TR symmetry.
Nevertheless, these two edge modes are Kramer doublets under the TR
transformation satisfying $T^2=-1$.
The celebrated Kane-Mele $Z_2$ argument for translational invariant systems
\cite{kane2005} can be generalized to these rotation invariant systems
by replacing linear momentum with the angular momentum.
If a given energy cuts odd numbers of helical edge modes, then any
TR invariant perturbation cannot mix these modes to open a gap.
Consequently, the topological nature of such a system is characterized by
the $Z_2$ index.

When loading fermions in the system, and if the Fermi energy cuts
the edge states, these helical edge states become active.
The effective helical edge Hamiltonian can be constructed by imposing an open
boundary at $r\approx l_T$.
The effective helical edge Hamiltonian in the basis of $j_z$ can
be written as
\bea
H_{edge}= \sum_{j_z}  (\frac{\hbar v_f}{l_T} |j_z|-\mu) \psi^\dagger_{n_r,j_z}
\psi_{n_r,j_z},
\label{eq:edge_1}
\eea
where $\mu$ is the chemical potential.
If the edge is considered locally flat, Eq. \ref{eq:edge_1} can be
rewritten in the plane-wave basis.
Due to the reflection symmetry with respect to the plane perpendicular
to the edge, the spin polarization for momentum $p$ along the
edge direction must lie in such a plane.
Also combining with TR symmetry, we have
\bea
H_{edge}&=& v [(\vec p \times \hat n) \cdot \hat z]
\Big ( \sigma_z \sin\eta +
(\vec \sigma \cdot \hat n) \cos\eta \Big),
\nn \\
\eea
where $\hat n$ is the local normal direction on the circular edge
in the 2D plane; $v$ is the linearized velocity of the edge modes
around Fermi energy;
$\eta$ is a parameter angle depending on the details of the systems.
There are the only terms allowed by rotation symmetry, TR symmetry, and
the vertical mirror symmetry in this system.
Each edge channel is a branch of helical one-dimensional Dirac fermion modes.

Parallel analysis can be applied to the helical surface states of the 3D
Hamiltonian Eq. (\ref{eq:single_3D}).
Again, due to the TR symmetry, surface states are helical instead of
chiral.
The topological class also belongs to $Z_2$.
If the surface is sufficiently large, and thus can be locally taken
as flat, we can construct the surface Dirac Hamiltonian around the Fermi
energy by using plane-wave basis basing on symmetry analysis.
First, due to the local $SO(2)$ rotational symmetry around $\hat n$,
the in-plane momentum $p_x$ and $p_y$ cannot couple to $\sigma_z$,
thus spin polarization for each in-plane momentum $(p_x,p_y)$ has
to lie in the $xy$-plane.
Generally speaking, the spin polarization vector $(s_x, s_y)$
form an angle $\eta$ with respect to $(p_x,p_y)$ and such
an angle is determined by the details of the surface.
Combine all the above information, we arrive at
\bea
H_{sfc}&=&v \Big\{ \sin \eta (\vec p \times \vec \sigma)
\cdot \hat n +\cos\eta \big[\vec p \cdot \vec \sigma
-(\vec p \cdot \hat n) (\vec \sigma \cdot \hat n)\big]
\Big\}, \nn \\
\label{eq:3D_surface}
\eea
where $\hat n$ is the local normal direction to the 2D surface.

\section{Topological spin textures and the quaternionic phase defects
in a harmonic trap}
\label{sect:texture}

In this section, we review the unconventional BECs with interactions
and SO couplings, including both Rashba and the 3D Weyl types,
in the harmonic trap.
The 2D Rashba case is presented in Sec. \ref{sect:2D_texture}.
The linear dependence on momentum of SO coupling invalidates the
proof of ``no-node'' theorem.
Consequently, a general feature of SO coupled BECs is the complex-valued
condensate wavefunctions and the spontaneous TR symmetry breaking.
For the Rashba case, the skyrmion type spin textures and half-quantum
vortex were  predicted in the harmonic trap \cite{Wu2011}.
Furthermore, due to the Landau level structures of single-particle states,
rich patterns of spin textures have been
extensively investigated in literatures
\cite{Wu2011,Hu2012a,Sinha2011}.
A nice introduction of topological defects in the ultra-cold
atom context can be found in Ref. [\onlinecite{Zhou2003}].


Even more interesting physics shows in 3D Weyl SO coupling, which will
be reviewed in Sec. \ref{sect:3D_texture}.
The non-trivial topology of the condensate wavefunction is most
clearly expressed in the quaternionic representation
\cite{Li2012,Kawakami2012}.
Quaternions are a natural extension of complex numbers as the first discovered
non-commutative division algebra, which has been widely applied in quantum
physics \cite{adler1995,finkelstein1962,balatsky1992}.
The condensation wavefunctions exhibit defects in the quaternionic
phase space as the 3D skyrmions, and the corresponding spin density
distributions are characterized by non-zero Hopf invariants.

\subsection{Half-quantum vortices and spin texture skyrmions with Rashba
SO coupling}
\label{sect:2D_texture}

Let us consider a 3D two-component boson system with contact spin-independent
interactions and with Rashba SO coupling in the $xy$-plane.
Since the Rashba SO coupling is 2D, interesting spin textures only distribute
in the $xy$-plane.
For simplicity, the condensate is set uniform along the $z$-direction,
then the problem is reduced to a 2D Gross-Pitaevskii (GP) equation as
\begin{eqnarray}
\Big\{
&-&\frac{\hbar^2  \nabla^2}{2M} +
i\hbar \lambda_R(\nabla_x \sigma_{y,\alpha\beta}-\nabla_y \sigma_{x,\alpha\beta})
+g n(r,\phi) \nonumber\\
&+&\frac{1}{2}M \omega^2 r^2 \Big\} \psi_\beta(r,\phi)
=E \psi_\alpha(r,\phi),
\label{eq:GPtrap}
\end{eqnarray}
where
$\psi_\alpha$'s with $\alpha=\uparrow,\downarrow$ are two-component
condensate wavefunctions; $n(r,\phi)$ is the particle density;
$g$ describes  the $s$-wave scattering
interaction.
The interaction energy scale is defined as $E_{int}=
g N_0/(\pi l_T^2 L_z)$, where $L_z$ is the system size
along the $z$-axis, and the dimensionless interaction parameter
is defined as $\beta=E_{int}/(\hbar \omega_T)$.

\begin{figure}
\centering\epsfig{file=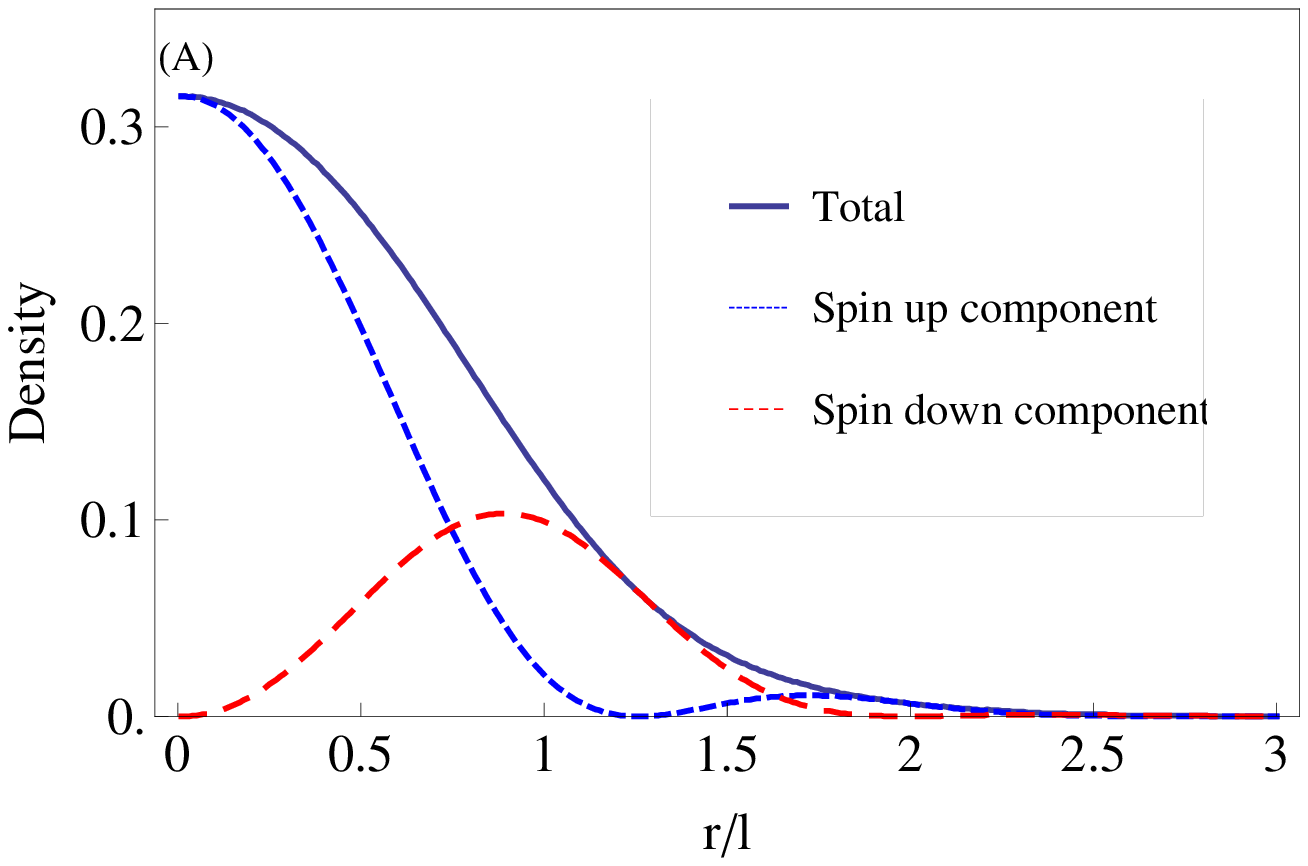,clip=1,width=\linewidth,
 angle=0}
\centering\epsfig{file=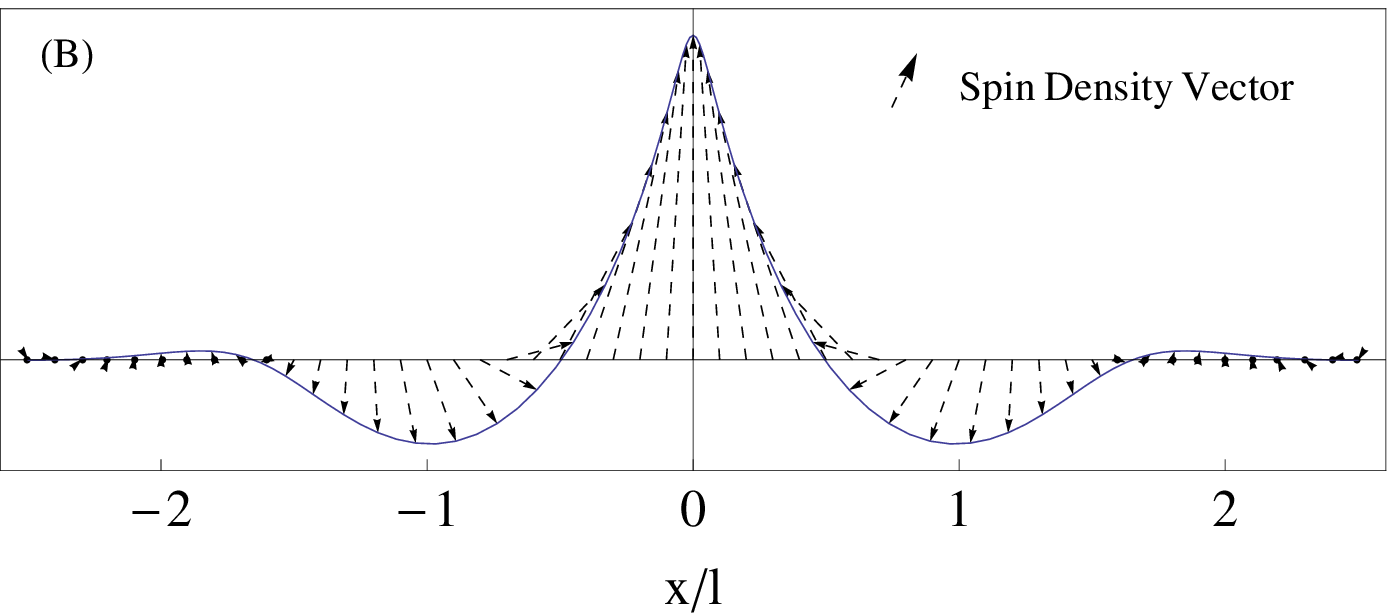,clip=1,width=\linewidth,
height=0.3\linewidth,angle=0}
\caption{(A) The radial density distribution of spin up and down
components, and the total density distribution in the unit of
$N_0=\int d^3\vec r \{|\psi_\uparrow(\vec r)|^2 + |\psi_\downarrow(\vec r)|^2\}$
at $\alpha=2$ and $\beta=5$.
(B) The skyrmion type spin texture configuration is plotted in the
$xz$-plane.
From Refs. \onlinecite{Wu2011}.
}
\label{fig:trap}
\end{figure}

We start with weak SO coupling, $\alpha \sim 1$, and with weak interactions.
In this case, the energy of single-particle ground state with
$j_z=\pm\frac{1}{2}$ is well separated from other states.
If interactions are not strong enough to mix the ground level with other
levels, the condensate wavefunction remains the same symmetry
structure carrying $j_z=\frac{1}{2}$, or, $-\frac{1}{2}$,
thus bosons condense into one of the TR doublets,
\bea
\psi_{\frac{1}{2}}(r,\phi)=\left( \begin{array}{c}
f(r) \\
g(r) e^{i\phi}
\end{array}
\right), ~~
\psi_{-\frac{1}{2}}(r,\phi)=\left(
\begin{array}{c}
-g(r) e^{-i\phi}\\
f(r)
\end{array} \right), \nn \\
\label{eq:TR_doublet}
\eea
where $f(r)$ and $g(r)$ are real radial functions.
In the non-interacting limit, $f(r)\approx J_0(k_{so} r)e^{-\frac{r^2}{4l_T^2}}
$ and $g(r)\approx J_1(k_{so} r) e^{-\frac{r^2}{4l_T^2}} $
as shown in Eq. (\ref{eq:2D_LL_rashba}).
Repulsive interactions expand the spatial distributions of $f(r)$ and
$g(r)$, but the qualitative picture remains.
Therefore, one spin component stays in the $s$-state and the other in
the $p$-state.
This is a half-quantum vortex configuration which spontaneously breaks
TR symmetry \cite{Wu2011}.

One possibility is that the condensate wavefunction may take linear
superpositions of the Kramer doublet in Eq. (\ref{eq:TR_doublet}).
The superposition principle usually does not apply due to the
non-linearity of the GP equation.
Nevertheless, if the interaction of the GP equation is spin-independent,
all the linear superpositions of the Kramer doublet in Eq.
(\ref{eq:TR_doublet})
are indeed degenerate.
This is an accidental degeneracy at the mean-field level which is not
protected.
Quantum fluctuations remove this degeneracy as shown in the exact
diagonalization calculation in Ref. \onlinecite{Hu2012a} and select
either one of $\psi_{\pm\frac{1}{2}}$.
In other words, quantum fluctuations can induce a spin-dependent
interaction beyond the mean-field level \cite{Wu2011}.
Certainly, we can also prepare the initial state with the average
$j_z$ per particle $\pm\frac{1}{2}$, say, by cooling down from the fully
polarized spin up or down state, then $\psi_{\pm\frac{1}{2}}$ will be reached.
On the other hand, if an additional spin-dependent interaction is
introduced,
\bea
H^\prime_{int}= g^\prime\int d^3 \vec r  \Big(n_\uparrow (r)-n_\downarrow (r)\Big)^2,
\eea
then even the mean-field level degeneracy is removed.
In this case, as shown in Ref. [\onlinecite{Ramachandhran2012}], the
condensate wavefunctions of $\psi_{\pm\frac{1}{2}}$ will also be selected.

The spin distribution of a condensate wavefunction is expressed as
\bea
\vec S(r,\phi)=\psi_\alpha^*(r,\phi)
\vec{\sigma}_{\alpha\beta} \psi_\beta(r,\phi),
\label{eq:hopf_1}
\eea
which is known as the 1st Hopf map.
Without loss of generality, the condensate of $\psi_{\frac{1}{2}}$
is considered, and its $\vec S(\vec r)$ is expressed as
\bea
S_x(r,\phi)&=& \rho \sin 2\gamma (r) \cos\phi, ~~
S_y(r,\phi)=   \rho \sin 2\gamma (r)\sin\phi, \nn \\
S_z(r,\phi)&=& \rho \cos 2\gamma (r),
\label{eq:spin}
\eea
where $\rho(r)=\sqrt{|f(r)|^2+|g(r)|^2}$, and the parameter angle $\gamma(r)$
is defined through
\bea
\cos \gamma(r)=\frac{f(r)}{\rho(r)}, \ \ \,
\sin\gamma(r)=\frac{g(r)}{\rho(r)}.
\eea
Since the Fourier components of $f(r)$ and $g(r)$ are located around
the Rashba ring in momentum space, they oscillate along the radial
direction with an approximated pitch value of $k_{so}$
as shown in Fig. \ref{fig:trap} (A).
Because $f(r)$ and $g(r)$ are of the $s$ and $p$-partial waves, respectively,
they are with a relative phase shift of $\frac{\pi}{2}$.
At $r=0$, $f(r)$ is at maximum and $g(r)$ is 0.
As $r$ increases, roughly speaking, the zero points of $f(r)$
corresponds to the extrema of $g(r)$ and vice versa, thus
$\gamma(r)$ spirals as $r$ increases.
At the $n$-th zero of $g(r)$ denoted $r_n$, $\gamma(r_n)=n\pi$
($n\ge 0$ and we define $r_0=0$).

Consequently, $\vec S$ spirals in the $zx$-plane along the $x$-axis as
shown in Fig. \ref{fig:trap} B.
The entire distribution of $\vec S$ can be obtained through a
rotation around the $z$-axis.
This is a skyrmion configuration which is a non-singular topological defect
mapping from the real space $R^2$ to the spin orientation space of the
$S^2$ sphere.
If the coordinate space is a closed manifold $S^2$, this mapping
is characterized by the integer
valued Pontryagin index $\pi_2(S^2)=Z$, or, the winding number.
However, the coordinate space is the open $R^2$, and $\rho(r)$ decays
exponentially at large distance $r\gg l_T$, thus the rigorously speaking
the covering number is not well-defined.
Nevertheless, in each concentric circle $r_n<r<r_{n+1}$,
$\gamma(r)$ varies from $n\pi$ to $(n+1)\pi$, which contributes
to the winding number by 1.
If we use the trap length scale $l_T$ as the system size,
the winding number is roughly at the order of $\alpha$.

The radial oscillation of the spin density is in analogy to the
Friedel oscillations in Fermi systems.
Around an impurity in electronic systems, the screening charge distribution exhibits the radial oscillation on top of the enveloping exponential decay.
The oscillation pitch is $2k_f$ reflecting the discontinuity of the Fermi
distribution on the spherical Fermi surface.
Different from the usual boson systems, the SO coupled ones have
a low energy ring structure in momentum space in analogous to the
Fermi surface, thus in real space spin density also oscillates
in the presence of spatial inhomogeneity.

\begin{figure}
\centering\epsfig{file=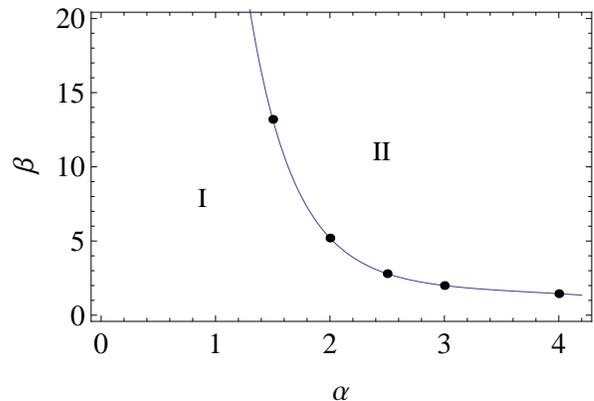,clip=1,width=0.9\linewidth}
\caption{The phase boundary of $\beta_c$ v.s $\alpha$ between
(I) the skyrmion condensates  with $j_z=\pm\frac{1}{2}$ and
(II) rotational symmetry breaking condensates.
From Ref. \onlinecite{Wu2011}.
}
\label{fig:transition}
\end{figure}

In the regime of intermediate SO coupling strength, the level spacing
between single particle states within the same Landau level is
suppressed as shown in Eq. (\ref{eq:2D_Rashba_gap}).
In the case that interactions are strong enough to mix
energy levels with different angular momenta in the same lowest Landau
level but not among different Landau levels, condensates do not keep
rotation symmetry any more.
The calculated phase boundary of interaction strength $\beta$ v.s.
SO coupling strength $\alpha$ is plotted in Fig. \ref{fig:transition}.
In this regime,
the distributions are no long concentric but split into multi-centers and
finally form a triangular skyrmion lattice structure as calculated
in Refs. [\onlinecite{Hu2012a}], [\onlinecite{Sinha2011}].
This 2D skyrmion lattice structure is a characteristic feature
brought by SO coupling.

\subsection{Plane-wave type condensations with Rashba SO coupling}
\label{sect:plane_wave}

If interactions are strong enough to mix states in different Landau levels,
then the influence of the confining trap is negligible.
The condensate configurations in the free space was calculated beyond
the mean-field GP equation level in Ref. [\onlinecite{Wu2011}].
Bosons select the superposition of a pair of states with opposite
momenta $\vec k$ and $-\vec k$ on the low energy Rashba ring to condense.
These spin eigenstates of these two states are orthogonal, thus the
condensate can avoid the positive exchange interactions.
As is well known, avoiding exchange energy is the main driving force
towards BEC.
For spin-independent interactions, the condensate wavefunctions
exhibit degeneracy at the Hartree-Fock level regardless of
the relative weight between these two plane-wave components.
This is a phenomenon of ``frustration''.
Quantum zero-point energy from the Bogoliubov quasi-particle spectra
selects an equal weight supposition through the ``order-from-disorder''
mechanism.
Such a condensate exhibits spin-spiral configuration.

Various literatures have also studied the case of spin-dependent
interactions in which the Hartree-Fock theory is already enough
to select either the spin-spiral state, or, a ferromagnetic
condensate with a single plane-wave component
\cite{Wang2010,Ho2011}.

\subsection{Quaternionic phase defects of the 3D Weyl SO coupling}
\label{sect:3D_texture}
Next we review the condensates with the 3D Weyl SO coupling in the harmonic
trap.
The corresponding GP equation is very similar to Eq. (\ref{eq:GPtrap}) of
the Rashba case.
Only slight modifications are needed by replacing the spatial dimension 2
with 3, and by replacing the Rashba term with $-i\hbar \lambda_W
\vec \nabla \cdot \vec \sigma$.
Amazingly, in this case condensate wavefunctions exhibit topological
structures in the quaternionic representation \cite{Li2012b}.

\begin{figure}
\centering\epsfig{file=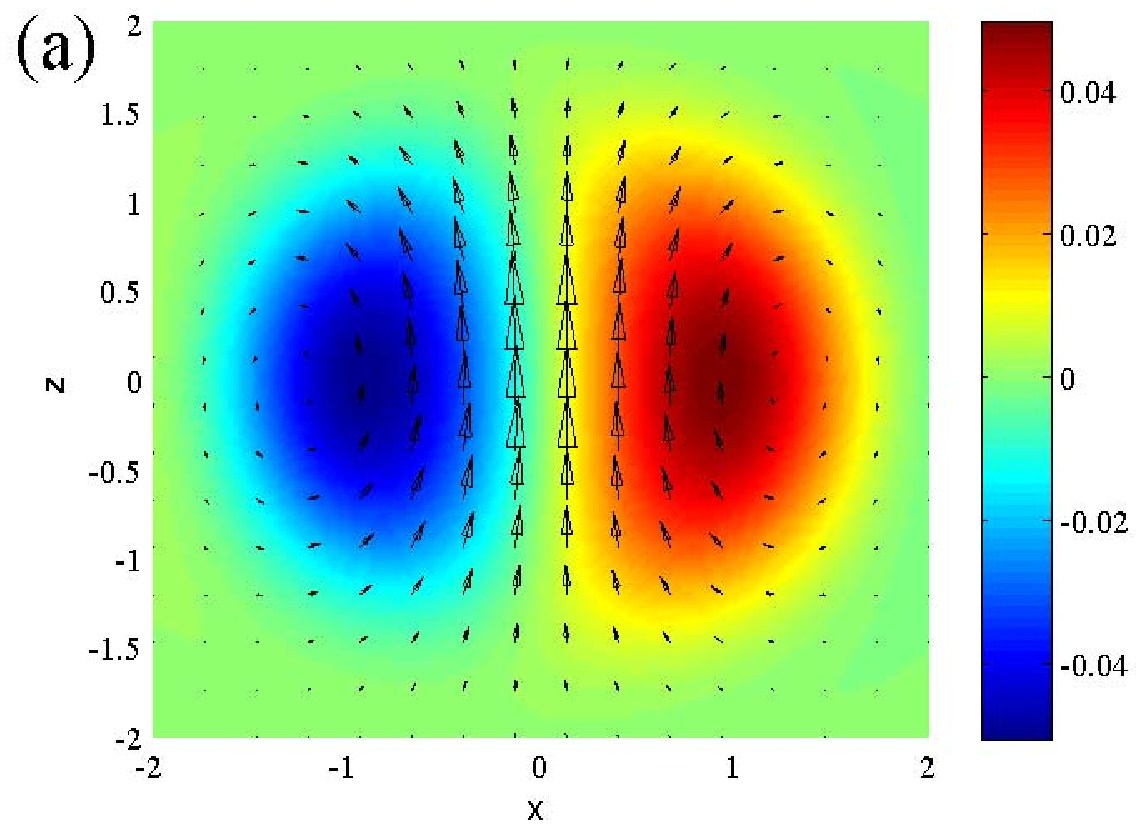,clip=1,width=0.8\linewidth}
\centering\epsfig{file=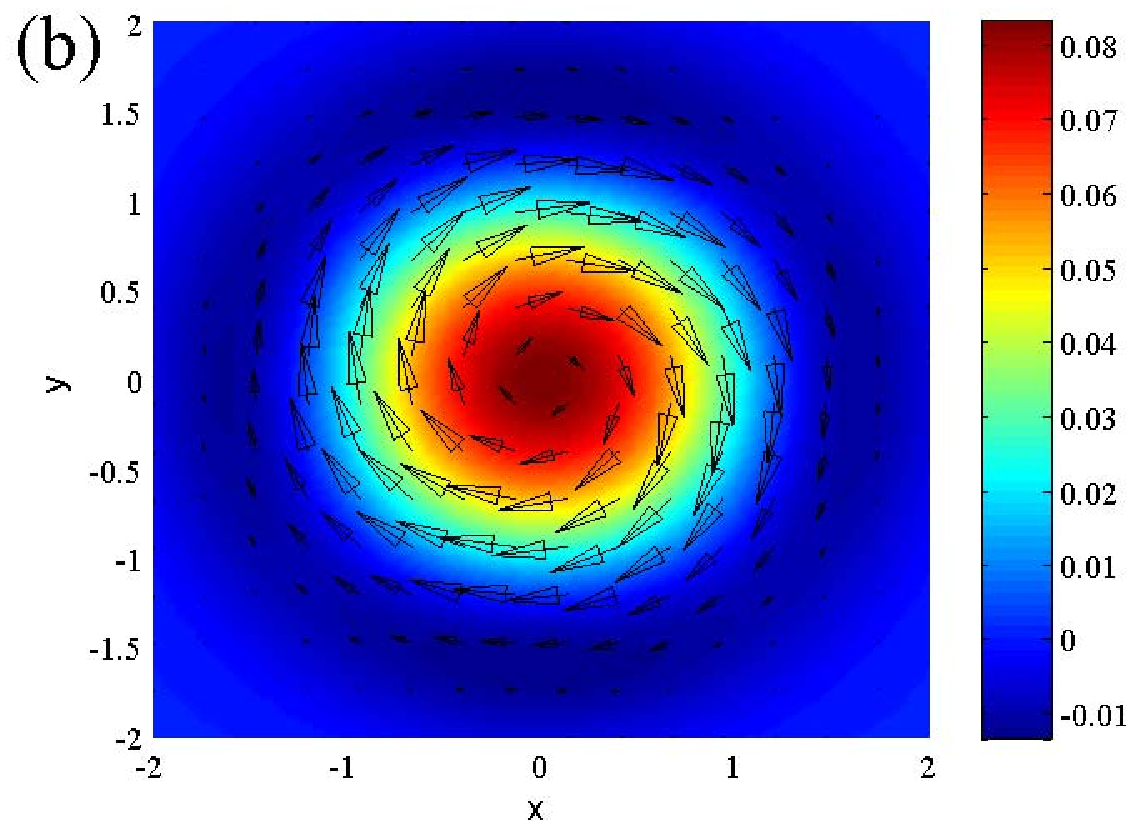,clip=1,width=0.8\linewidth}
\centering\epsfig{file=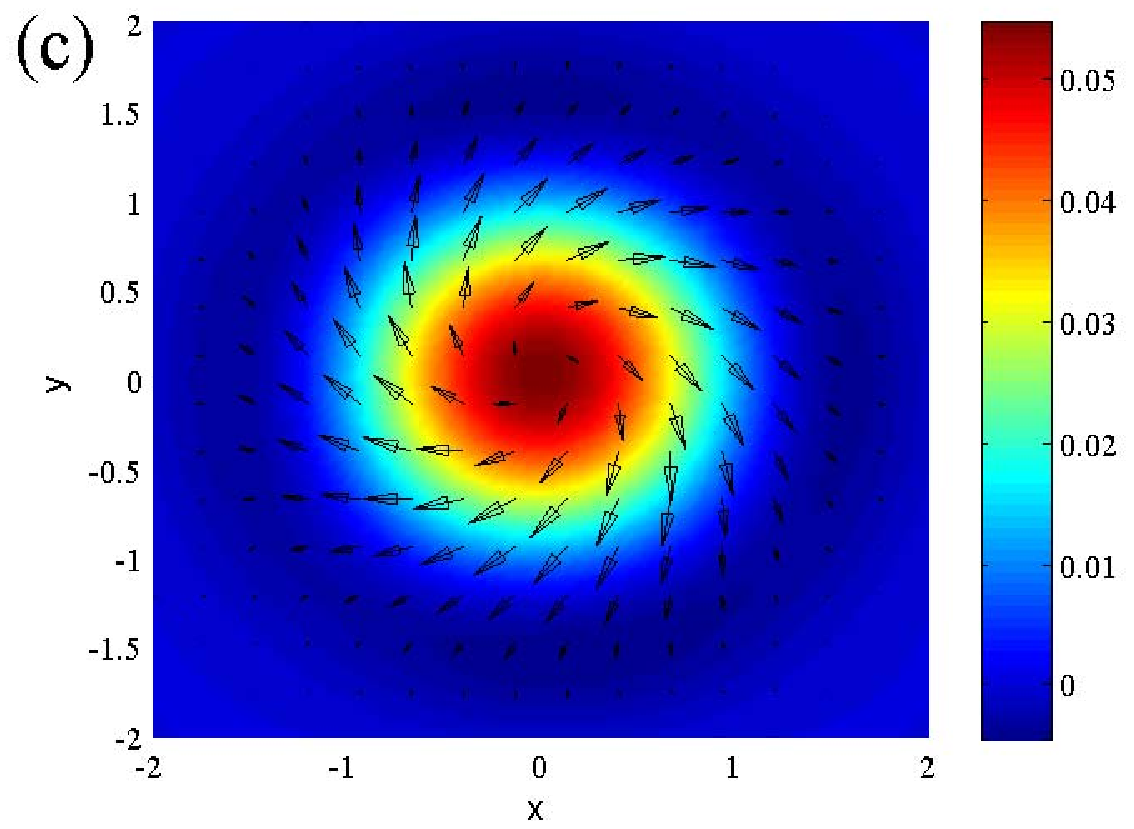,clip=1,width=0.8\linewidth}
\caption{The distribution of $\vec{S} (\vec r)$ in a) the $xz$-plane
and in the horizontal planes with b) $z=0$ and
c) $z/l_T=\frac{1}{2}$ with $\alpha=1.5$, $c=1$, and $\beta=30$.
The color scale shows the magnitude of out-plane component
$S_y$ in a) and $S_z$ in b) and c).
The 3D distribution of $\vec S(\vec r)$ is topologically non-trivial
with a non-zero Hopf invariant.
The length unit in all the figures is $l_T$.
From Ref. \onlinecite{Li2012b}.
}
\label{fig:spin}
\end{figure}

\subsubsection{The quaternionic representation}
Just like a pair of real numbers form a complex number, the two-component
spinor $\psi=(\psi_\uparrow,\psi_\downarrow)^T$ is mapped to a single
quaternion following the rule
\bea
\xi =\xi_0 +\xi_1 i + \xi_2 j +\xi_3 k,
\eea
where $\xi_0=\Re\psi_\uparrow, \xi_1=\Im \psi_\downarrow, \xi_2=-\Re\psi_\downarrow,
\xi_3=\Im\psi_\uparrow$.
$i,j,k$ are the imaginary units satisfying $i^2=j^2=k^2=ijk=-1$,
and the anti-commutation relation $ij=-ji=k$.
Quaternion can also be expressed in the exponential form as
\bea
\xi=|\xi| e^{\omega \gamma}
=|\xi|(\cos\gamma +\omega\sin \gamma),
\eea
where $|\xi|=\sqrt{\xi_0^2+ |\vec \xi|^2 }$ and
$|\vec \xi|^2=\xi_1^2+\xi_2^2+\xi_3^2$;
$\omega$ is the unit imaginary unit defined as
$\omega=(\xi_1 i +\xi_2 j + \xi_3 k)/|\vec \xi|$
which satisfies $\omega^2 =-1$;
the argument angle $\gamma$ is defined as
$\cos\gamma=\xi_0/|\xi|$ and $\sin\gamma=|\vec \xi|/|\xi|$.

Similarly to the complex phase $e^{i\phi}$ which spans a unit circle,
the quaternionic phases $e^{\omega \gamma}$ span a unit three dimensional
sphere $S^3$.
The spin orientations lie in the $S^2$ Bloch sphere.
For a quaternionic wavefunction, its corresponding spin distribution
is defined through the 1st Hopf map defined in  Eq.
(\ref{eq:hopf_1})
as a mapping $S^3\rightarrow S^2$.
Due to the homotopy groups\cite{wilczek1983,nakahara2003}
$\pi_3(S^3)=Z$ and $\pi_3(S^2)=Z$, both quaternionic condensate
wavefunctions and spin distributions can be non-trivial.
The winding number of $S^3\rightarrow S^3$ is the 3D skyrmion number,
and that of the $S^3\rightarrow S^2$ is the Hopf invariant, both are
integer-valued.

Let us apply the above analysis to  the lowest single-particle
level with $j_z=j=\frac{1}{2}$
\bea
\psi_{j=j_z=\frac{1}{2}}(r,\hat \Omega)=f(r) Y_{+,\frac{1}{2},0,\frac{1}{2}}
(\hat\Omega) + ig(r) Y_{-,\frac{1}{2},1,\frac{1}{2}}(\hat\Omega),
\nn \\
\label{eq:3D_cond}
\eea
where $Y_{+,\frac{1}{2},0,\frac{1}{2}}(\hat\Omega)=(1,0)^T$
and $Y_{-,\frac{1}{2},1,\frac{1}{2}}(\hat\Omega)
=(\cos\theta,\sin\theta e^{i\phi})^T$.
As shown in Eq. (\ref{eq:3D_WF}), in the non-interacting limit,
$f(r)\approx j_0(k_{so} r)e^{-\frac{r^2}{4l_T^2}}
$ and $g(r)\approx j_1(k_{so} r) e^{-\frac{r^2}{4l_T^2}}$.
The corresponding quaternionic expression is
\bea
\xi_{j=j_z=\frac{1}{2}}(r, \hat \Omega)
= \rho(r) e^{ \omega(\hat\Omega) \gamma(r)},
\label{eq:qua_phase}
\eea
where $\rho(r)$ and $\gamma(r)$ are defined in the same way as
the 2D case in Eq. (\ref{eq:spin});
the imaginary unit,
\bea
\omega(\hat\Omega)=
\sin\theta \cos\phi ~i +\sin\theta \sin\phi~ j + \cos\theta ~ k,
\eea
is along the direction of
$\hat\Omega$.

\subsubsection{The skyrmion type 3D quaternionic phase defects }

The analysis on the topology of the Weyl condensates can be performed in
parallel to the above 2D Rashba case.
Again in the case of weak SO coupling, interactions only expand
the spatial distribution of $f(r)$ and $g(r)$ in Eq.
(\ref{eq:3D_cond})
from their non-interacting forms.
The radial wavefunction $f(r)$ and $g(r)$ follow the same oscillating
pattern as those in the Rashba case, thus so does the parameter angle
$\gamma(r)$ which starts from 0 at the origin and reaches $n\pi$
at the $n$-th zero of $f(r_n)=0$.
For the quaternionic phase $e^{\omega(\hat\Omega)\gamma(r)}$, its
imaginary unit $\omega(\hat \Omega)$ is of one-to-one correspondence
to every direction in 3D, thus it
exhibits a non-trivial mapping from the 3D coordinate space $R^3$
to $S^3$, which is known as a 3D skyrmion configuration.

For a closed 3-manifold, the Pontryagin index of winding number is
$\pi_3(S^3)=Z$, i.e., integer.
Again here the real space is open.
In each concentric spherical shell with $r_n< r <r_{n+1}$ whose
thickness is at the scale of $l_{so}$, $\gamma(r)$
spirals from $n\pi$ to $(n+1)\pi$, thus this shell contributes 1 to
the winding number from real space to the quaternionic phase manifold.
If the system size is truncated at the trap length $l_T$, again
the winding number is approximately $\alpha$.
In comparison, in the 2D Rashba case reviewed in Sec. \ref{sect:2D_texture},
$\vec S(\vec r)$ exhibits the 2D skyrmion configuration
\cite{Wu2011,Hu2012a,Sinha2011}, but
condensation wavefunctions have no well-defined topology due
to the fact that $\pi_2(S^3)=0$.

A comparison can be made with the $U(1)$ vortex in the single-component BEC.
In 2D, it is a topological defect with a singular core.
Moving around the circle enclosing the core, the phase winds
from $0$ to $2\pi$, and thus the winding number is 1.
The above 3D skyrmion phase defect is a natural generalization to the
two-component case whose phase space is $S^3$ in the quaternionic
representation and is isomorphic to the SU(2) group manifold.
These 3D skyrmions are non-singular defects similar to a
1D ring of rotating BEC which carries a non-zero phase winding number
but the vortex core lies outside the ring.

\begin{figure}
\includegraphics[width=0.8\linewidth]{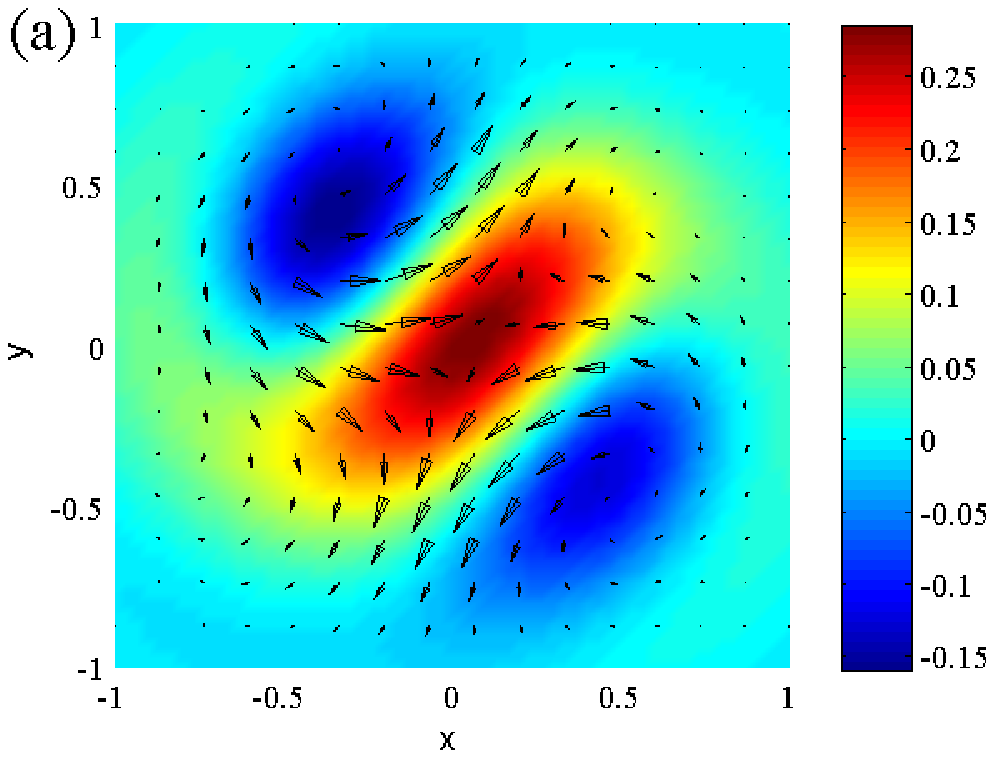}
\includegraphics[width=0.8\linewidth]{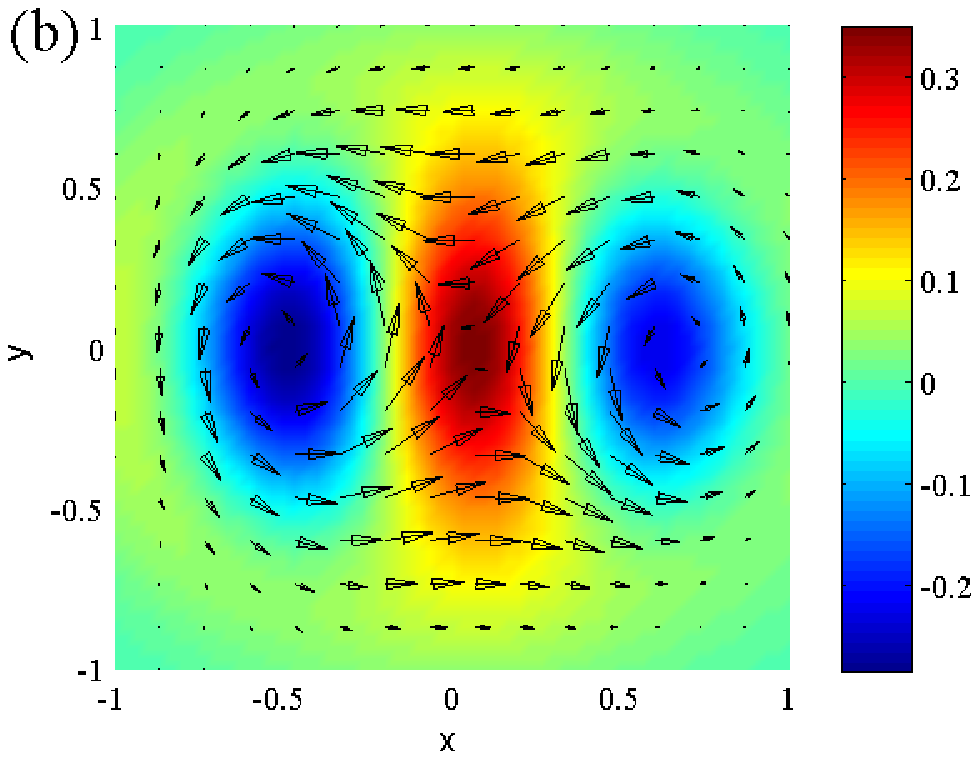}
\includegraphics[width=0.8\linewidth]{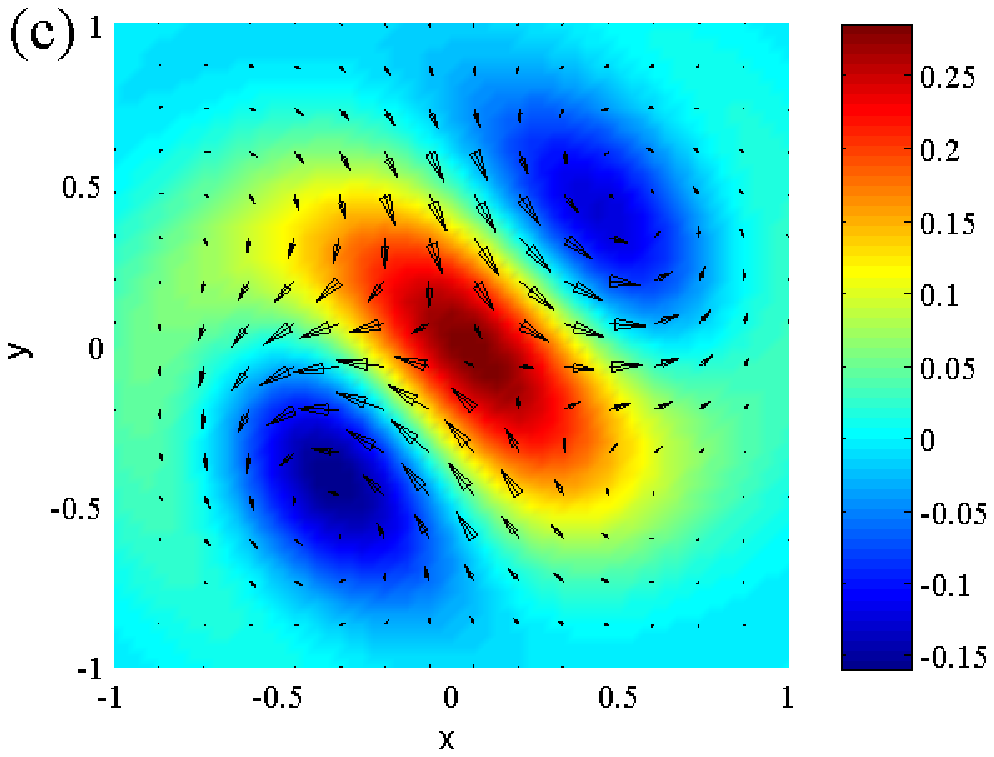}
\caption{The distribution of $\vec S(\vec r)$ in horizontal cross-sections
with a) $z/l_T=-0.5$, b) $z/l_T=0$, c) $z/l_T=0.5$, respectively.
The color scale shows the value of $S_z$, and parameter values
are $\alpha=4$, $\beta=2$, and $c=1$.
}
\label{fig:spin_2}
\end{figure}

\subsubsection{Spin textures with non-zero Hopf invariants}

The non-trivial topology of the condensate wavefunctions leads to a
topologically non-trivial distribution of spin density $\vec S(\vec r)$.
The 1st Hopf map defined in Eq. (\ref{eq:hopf_1}) becomes very elegant
in the quaternionic representation as
\bea
S_x i + S_y j + S_z k=\frac{1}{2}\bar \xi k \xi,
\label{eq:hopf}
\eea
where $\bar \xi=\xi_0 - \xi_1 i - \xi_2 j - \xi_3 k$ is the quaternionic
conjugate of $\xi$.
For the condensate wavefunction Eq. (\ref{eq:qua_phase}),
$\vec S(\vec r)$ is calculated as
\bea
\left[
\begin{array}{c}
S_x(\vec r)\\
S_y(\vec r)
\end{array}
\right] &=&g(r) \sin\theta
\left[
\begin{array}{cc}
\cos\phi& -\sin\phi\\
\sin\phi& \cos\phi
\end{array}
\right] \left[
\begin{array}{c}
g(r)\cos\theta\\
f(r)
\end{array}
\right], \nn \\
S_z(\vec r)&=& f^2(r)+g^2(r) \cos2\theta,
\eea
which exhibits a perfect axial symmetry around the $z$-axis.
$\vec S(\vec r)$ is plotted in Fig. \ref{fig:spin} at
different cross sections.
In the $xy$-plane, it exhibits a 2D skyrmion pattern,
whose in-plane components are along the tangential direction.
As the horizontal cross-section shifted along the $z$-axis,
$\vec S(\vec r)$ remains the 2D skyrmion-like, but its in-plane
components are twisted around the $z$-axis.
According to the sign of the interception $z_0$, the twist
is clockwise or anti-clockwise, respectively.
This 3D distribution pattern of $\vec S(\vec r)$ is characterized
by an integer valued Hopf invariant characterized by $\pi_3(S^2)=Z$.

As SO coupling strength increases, condensates break rotational symmetry
by mixing different states with different values of $j$ in the lowest
Landau level.
Even at intermediate  level of SO coupling, rich patterns appear.
The quaternionic phase defects and the corresponding spin textures
split into multi-centered pattern as plotted in Fig. \ref{fig:spin_2}
for different horizontal cross-sections.
In the $xy$-plane, $\vec S$ exhibits a multiple skyrmion configuration
as shown in the combined pattern of the in-plane and z-components.
Again this pattern is twisted by rotating around the $z$-axis as
the interception value $z$ varies.
Thus the whole 3D configuration also possesses non-trivial
Hopf invariant.

In particular, in the case of $\alpha\gg 1$, it is expected that a 3D lattice structure of
topological defects maybe formed,  which
is a generalization of the 2D skyrmion lattice configuration
in Refs. [\onlinecite{Hu2012a}] into 3D.
Again, if interaction is very strong to mix states in different
Landau levels, the condensate will become plane-wave-like or
superpositions of SO coupled plane-waves
\cite{Li2012b}.


\section{Vortex configurations of SO coupled BECs in a rotating trap}
\label{sect:vort}
Next we review the vortex configurations of SO coupled unconventional BECs
in rotating traps.
From a more general framework, the above-considered SO coupling can be
viewed as particles subject to non-Abelian gauge fields.
On the other hand, the Coriolis force from rotation behaves as an
effective Abelian vector potential.
Therefore, in a rotating trap, atom-laser coupling provides an elegant
way to study the effects of these two different effective gauge fields.
We only consider the rotating systems with the Rashba SO coupling.

\subsection{Hamiltonians of SO coupled bosons in a rotating trap}

\begin{figure}
\centering\epsfig{file=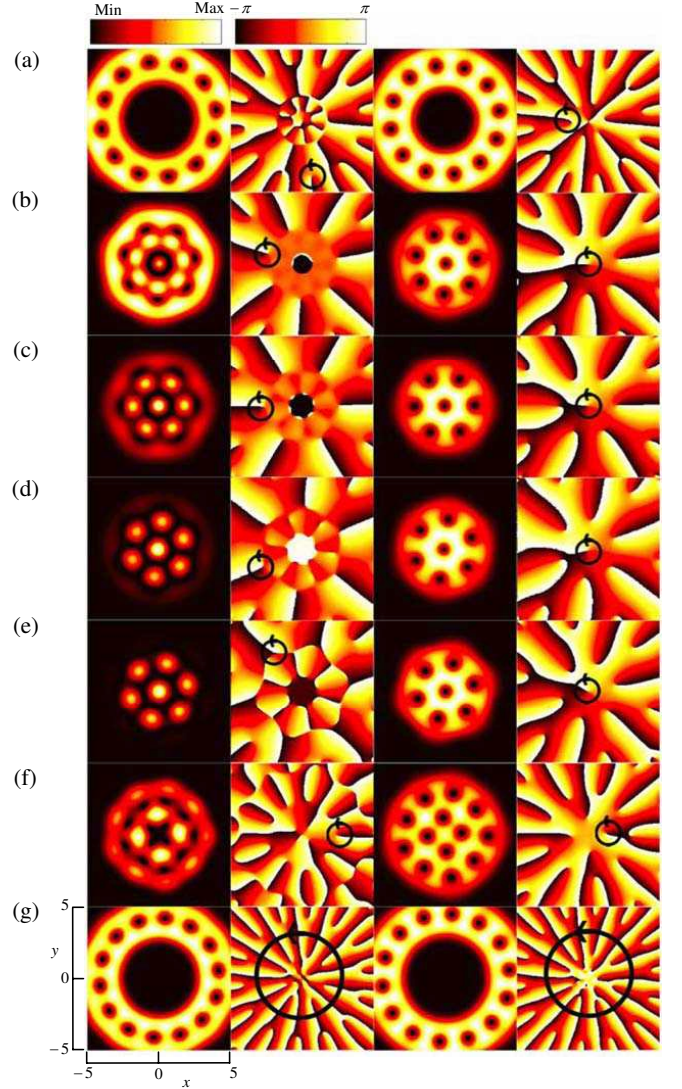,clip=1,width=\linewidth}
\caption{
From left to right: the density and phase profiles of $\psi_\uparrow(r)$
and $\psi_\downarrow(r)$ with parameter values of $\alpha=0.5$,
$\beta=10$, $\rho=0.97$.
From $(a)$-$(g)$, $\gamma$ is taken as $0.5$, $0.25$, $0.1$, $0.0$,
$-0.1$, $-0.3$, and $-0.5$, respectively.
At small values of $|\gamma|$ in [$c, d$, and  $e$], a skyrmion lattice  is formed
near the trap center.
By increasing the magnitude of  $|\gamma|$ [ $(b)$ and $(f)$], the skyrmion
lattice evolves to the normal vortex lattice.
For the large value of $|\gamma| = 0.5$ [$(a)$ and
$(g)$], the condensates show a lattice configuration around a ring. The
black circle with an arrow indicates the direction of the circulation
around the vortex core. The unit of length for the figures is $l_T$.
From Ref. [\onlinecite{ZhouXF2011}].
}
\label{fig:vort_latt}
\end{figure}

Ultracold atoms in a rotating trap share similar physics of electrons subject
to magnetic fields due to the similarity between Lorentz and Coriolis forces.
Depending on the experimental implementations of rotation,
Hamiltonians can be of different types \cite{ZhouXF2011,Radic2011,Xu2011}.
As pointed out in Ref. [\onlinecite{Radic2011}], because the current
experiment setup breaks rotation symmetry, rotating SO coupled
BECs is time-dependent in the rotating frame, which is a
considerably more complicated problem than the usual rotating BECs.
Nevertheless, below we only consider the situation of the isotropic Rashba
SO coupling, such that it in principle can be implemented
as a time-independent problem in the rotating frame.

The effect of rotation should be described by the standard
minimal substitution method as presented in Ref. [\onlinecite{ZhouXF2011}].
The non-interacting part of the Hamiltonian is
\bea
H_0 &=& \int d^3 \vec{r} \psi_{\mu}^{\dag}(\vec{r})
\big [ \frac{1}{2 M} (-i \hbar \vec{\nabla} + M \lambda
\hat z \times \vec \sigma -
\vec{A})^2 -\mu
\label{eq:h0}
\nn \\
&+&V_{tr}(\vec{r})- \frac{1}{2}M\Omega_z^2 (x^2+y^2) \big]_{\mu \nu}
\psi_{\nu}(\vec{r}),
\label{eq:rotation}
\eea
where $\vec{A}=m\Omega_z \vec{r} \times \hat{z}$, and the last term is the centrifugal potential due to rotation.

Note that due to the presence of SO coupling, we should carefully
distinguish the difference between mechanical and canonical
angular momenta.
The mechanical one should be defined according to the minimal substitution as
\bea
L^{mech}=L_z+M\lambda (x \sigma_x +y\sigma_y),
\eea
where $L_z$ is the usual canonical angular momentum.
Expanding Eq. (\ref{eq:rotation}), it is equivalent to Eq.
(\ref{eq:single_rashba}) plus the term of angular velocity
$\Omega_z$ coupling to $L^{mech}$ as
\bea
H_{rot}= - \Omega_z \int d^3 \vec{r} \psi_{\mu}^{\dag}(\vec{r})
\Big[ L^{mech} \Big ]_{\mu\nu} \psi_{\nu}(\vec{r}).
\eea
Thus in the rotating frame, the effect of $\Omega_z$ is not only just
$\Omega_z L_z$ as usual, but also an extra effective radial Zeeman
term as
\bea
\vec{B}_{R}(\vec r) = \Omega_z M \lambda \vec r.
\label{eq:extra}
\eea
Such a term is often missed in literatures.
As will be shown below, it affects the ground state vortex configurations
significantly and thus should not be overlooked.

To make the model more adjustable, an external spatially dependent
Zeeman field $\vec B_{ex} = B\vec r$ is intentionally introduced  as
\bea
H_B&=& - B \int d^3r ~\psi^\dagger_\mu(\vec r) ~(x \sigma_x + y
\sigma_y)_{\mu\nu} ~\psi_\nu (\vec r), ~~
\label{eq:zeeman}
\eea
which shares the same form as  Eq. (\ref{eq:extra}).
Experimentally, such a Zeeman field can be generated through coupling two
spin components using two standing waves in the $x$ and $y$-directions
with a phase difference of $\frac{\pi}{2}$.
The corresponding Rabi coupling is
\bea
-\Omega^\prime \Big\{\sin(k_L x) +i \sin(k_L y) \Big\}
\psi^\dagger_\downarrow(\vec r) \psi_\uparrow(\vec r) + h.c.
\eea
In the region of $|x|, |y|\ll 2\pi/k_L$, it reduces to the desired form
of Eq. (\ref{eq:zeeman}) with $B=\Omega^\prime k_L$.
Such a term compensates the non-canonical part of the mechanical momentum
in $H_{rot}$, which renders the model adjustability in a wider range.

\subsection{SO coupled bosons in rotating traps}
Now we turn on interactions and obtain the ground state condensate
numerically by solving the SO coupled GP equations which have been
reduced into the dimensionless form as
\bea
\label{eq:GP}
\mu \tilde{\psi}_{\uparrow}&=&\hat{T}_{\uparrow \nu}
\tilde{\psi}_{\nu} + \beta ( |\tilde{\psi}_{\uparrow}|^2
+ |\tilde{\psi}_{\downarrow}|^2 )
\tilde{\psi}_{\uparrow}, \nn \\
\mu \tilde{\psi}_{\downarrow}&=&\hat{T}_{\downarrow \nu}
\tilde{\psi}_{\nu} + \beta ( |\tilde{\psi}_{\downarrow}|^2 +
|\tilde{\psi}_{\uparrow}|^2 )
\tilde{\psi}_{\downarrow},
\eea
where $(\tilde\psi_\uparrow, \tilde\psi_\downarrow)$ are normalized
according to the condition
$\int d\vec r^2 (|\psi_\uparrow|^2+|\psi_\downarrow|^2)=1$.
$\hat T$ is defined as
\bea
\hat{T}&=&-\frac{1}{2}l^2_T(\partial^2_x + \partial^2_y) +
\alpha l_T (-i \partial_y \sigma_x + i \partial_x \sigma_y) \nn \\
&+& \frac{1}{2 l_T^2}(x^2+y^2) -
\frac{\rho}{l_T} (-i x\partial_y +iy\partial_x) \nn \\
&-& \alpha \frac{\kappa}{l_T} (x \sigma_x + y
\sigma_y),
\eea
where $\rho=\Omega_z/\omega$; $\kappa=\gamma+\rho$;
and $\gamma = B/(M \omega \lambda)$ is defined for the extra radial
Zeeman field in Eq. (\ref{eq:zeeman}).

\subsubsection{The skyrmion lattice structure in the weak SO coupling}

\begin{figure}
\centering\epsfig{file=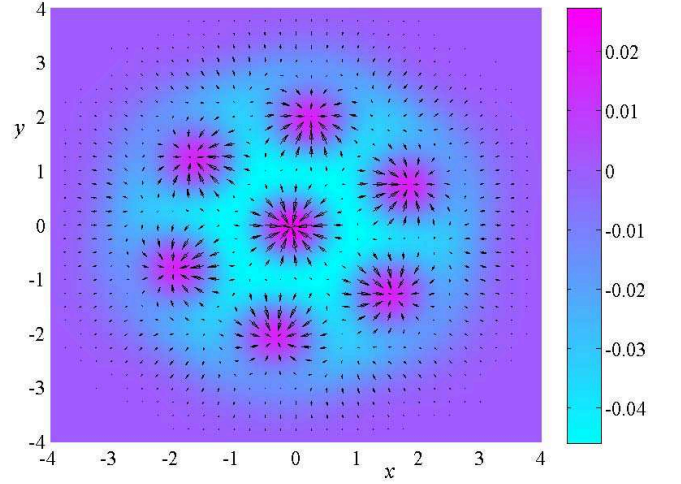,clip=1,width=\linewidth}
\caption{
The spin density distribution of the condensate of Fig. \ref{fig:vort_latt}
(d).
The projection of $\langle \vec{\sigma} \rangle$ in the $xy$-plane
is shown as black vectors. A color map is used to illustrate the
$\langle \sigma_z \rangle$ component.
The unit of length for the figure is $l_T$.
From Ref. [\onlinecite{ZhouXF2011}].
}
\label{fig:skymion}
\end{figure}

\begin{figure}[!htbp]
\centering\epsfig{file=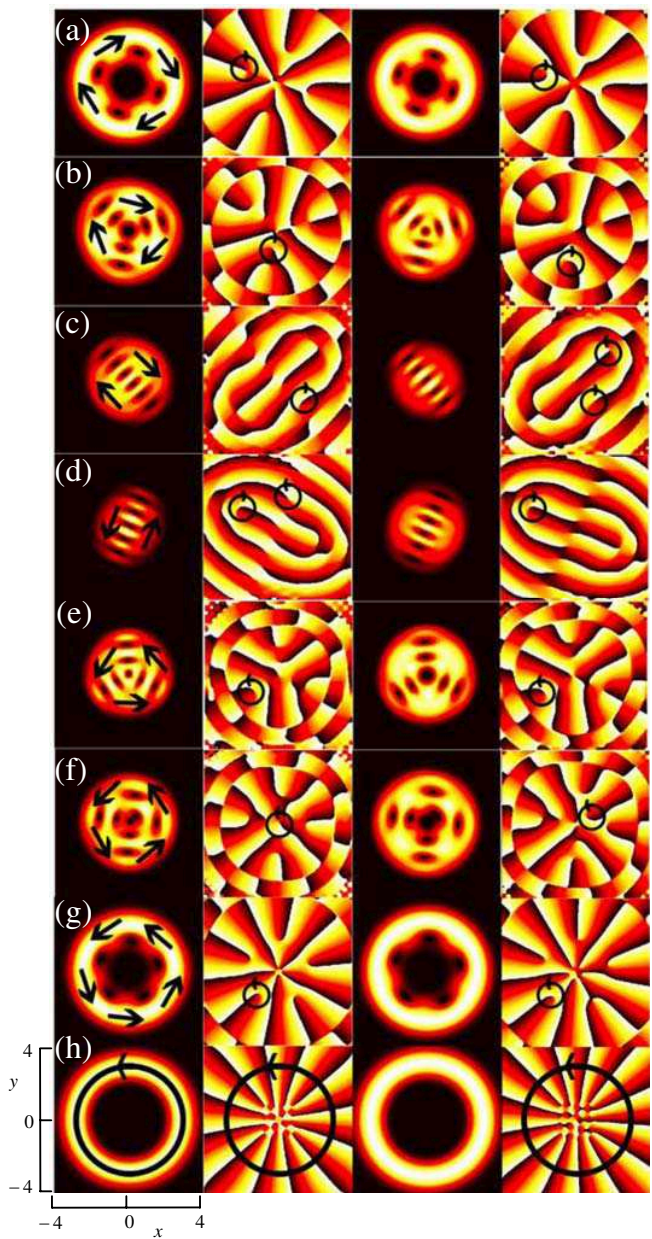,clip=1,width=\linewidth}
\caption{From left to right: the density and phase profiles of
$\psi_\uparrow(r)$ and $\psi_\downarrow(r)$ with parameter
values of $\alpha=4$, $\beta=20$, $c=1$, and $\rho=0.1$.
From $(a)$-$(h)$, $\gamma$ is taken as $0.5$, $0.3$, $0.1$, $-0.05$,
$-0.25$, $-0.35$, $-0.6$, and $-0.7$, respectively.
The black arrow in each domain represents the local wavevector direction
of the corresponding plane-wave state, which shows a clockwise or
counter-clockwise configuration depending on the sign of $\gamma$.
For sufficiently large values of $|\gamma|$, condensates distribute
around a ring in space forming a giant vortex.
The color scales for the density and phase distributions are
the same as that in Fig. \ref{fig:vort_latt}.
The black circle with an arrow indicates the the direction
of the circulation around the vortex core.
The unit of length for the figures is $l_T$.
From Ref. [\onlinecite{ZhouXF2011}].}
\label{fig:domain}
\end{figure}

Rich structures of vortex lattices appear in the case of
weak SO coupling.
In Fig. \ref{fig:vort_latt}, the SO parameter is taken as $\alpha=0.5$.
Both the density and phase patterns for $\psi_\uparrow(r)$ and
$\psi_\downarrow(r)$ are depicted.
Let us first consider the case of pure rotation of $B=0$ as shown in Fig.
\ref{fig:vort_latt} ($d$), i.e., $\gamma=0$.
For $\psi_\uparrow(r)$, its density distribution exhibits several
disconnected peaks.
By contrast, the usual vortex lattices show disconnected
low density vortex cores.
The phase distribution $\psi_\uparrow(r)$ exhibits singular points
around which phases wind 2$\pi$.
These singular points are squeezed out to the low density region near the edge.
For $\psi_\downarrow(r)$, its vortex cores are pinned by peaks of the density of $\psi_\uparrow(r)$.
Combine the distributions of $\psi_\uparrow(r)$ and $\psi_\downarrow(r)$
together, the condensate exhibits a skyrmion lattice configuration
with the spin distribution  $\langle \vec S \rangle$
shown in Fig. \ref{fig:skymion}.

Turning on the external Zeeman field $\vec B_{ex}$ of Eq.
(\ref{eq:zeeman})
changes the lattice configuration.
If $\vec B_{ex}\parallel \vec B_R$, the parameter $\gamma>0$,
and otherwise $\gamma<0$.
For both $\gamma>0$ and $\gamma<0$, if $|\gamma|$ is small,
the skyrmion lattice structures remain
as depicted in Fig. \ref{fig:vort_latt} ($b$, $c$, $e$, $f$).
Increasing $|\gamma|$ further, the condensates of both
spin components are pushed outwards,
and distribute around a ring with a giant vortex core, as shown in
Fig. \ref{fig:vort_latt} ($a$) and ($g$).
This ring is the location of potential minima shifted from the trap
center by the $H_B$ term.
In all cases in Fig. \ref{fig:vort_latt} ($a$-$g$),
the difference of vortex numbers between the spin-up and down
components is one.
This is a characteristic feature brought by the Rashba
SO coupling.

\subsubsection{The domain wall structure in the strong SO coupling}
As discussed in section \ref{sect:plane_wave}, in the case of strong SO
coupling, if interactions are also strong, condensates are nearly
suppositions of SO coupled plane-wave states subject to the trap
boundary condition.
We consider the effect of rotation in this case.
Limited by the numeric convergence, only a small rotation
angular velocity is considered.
The term of the external Zeeman field $\vec B_{ex}$ of
Eq. (\ref{eq:zeeman}) is also applied, which enriches the structures
of the condensates as shown in Fig. \ref{fig:domain}.

The characteristic feature is that the condensate around the trap center
is broken into several domains.
Inside each domain, the condensate is approximated as a plane-wave state.
Wavevectors are arranged such that the local spin polarizations are
parallel to $\vec{B}_{ex}$ in order to minimize the Zeeman energy.
At small values of $|\gamma|$ as shown in Fig. \ref{fig:domain} ($c$,
and $d$), a line of vortices appear at the boundary to separate two
adjacent domains.
Depending on the direction of $\vec B_{ex}$, the wavevectors inside
domains can be clockwise or counterclockwise.
For instance, when $\gamma>0$, a clockwise configuration of these
local wavevectors is selected in the condensate.
As further increasing $|\gamma|$, different domains connect together
to form a giant vortex as shown in Fig. \ref{fig:domain} ($a$, $g$, $h$).
Both spin components overlap with each other, and distribute
around a ring with the radius of $\alpha|\gamma|l_T$.
The spin textures lie along the radial direction to minimize
the magnetic energy.


\section{Strongly correlated phases of SO coupled bosons}
\label{sect:Strong}

Ultracold atoms with SO coupling provide us an unique opportunity of
manipulating
strongly correlated topological states in a highly controllable way.
With current technology, it becomes experimental feasible to
implement effective magnetic fields in atomic systems through
laser-atom interactions \cite{Lin2009,struck2012,jimenez2012}.
Since these systems share similar Hamiltonians with the solid state
quantum Hall physics, this enables us to investigate
various strongly correlated physics in the presence of strong interactions.
In addition, the realization of SO coupling effects using atoms also paves the way of
searching for novel phases with non-trivial topology beyond
traditional electronic systems.

Considerable progress has also been made along this direction.
For instance, in Ref. [\onlinecite{Ramachandhran2013}], the authors
have calculated the few-body physics using exact diagonalization scheme
for $2$D SO coupled bosons confined in a harmonic trap.
A strongly correlated ground state with non-trivial topology has been
reported at strong interaction strengths.
For pseudospin-$\frac{1}{2}$ bosons subject to effective magnetic
fields perpendicular to the $2$D plane with periodic boundary
condition, it has been shown in Ref. [\onlinecite{Burrello2010,Burrello2011}]
that a Rashba-like SO coupling favors different quantum Hall
phases depending on the coupling strengths.
When the lowest Landau level approximation is valid for large Landau level gaps,
the spin-polarized fractional quantum Hall states are formed for short-range interactions with Abelian excitations.
Around some particular degenerate points where two Landau levels have the same energy, the ground
states of the system are defined as deformed Halperin states.
The non-Abelian nature of their anyonic excitations is crucial for
the realization of topological quantum computations \cite{kitaev2003, Nayak2008}.
Numerical investigations of these strongly correlated states at different
filling factors have also been addressed recently \cite{grass2012b,grass2012c,palmer2011,komineas2012}.
We note that due to the rapid development of the area, many novel topological
quantum phases are expected to be found within SO coupled atomic systems.


\section{Magnetic phases of SO coupled bosons in optical lattices}
\label{sect:Mott}
It is natural to further consider the SO coupling effect in optical lattices,
in particular in the Mott-insulating states
\cite{Cai2012,Cole2012,Radic2012,Gong2012,Mandal2012}.
Due to SO coupling, hopping amplitudes are spin-dependent whose values
vary non-monotonically as increasing SO coupling strength.
The spin-dependent hopping leads to the Dzyaloshinsky-Moriya (DM) type
spin exchange models in the Mott-insulating states, which results
in rich spin ordering patterns.

Due to the scope and limitation of this article, we do not cover the synthetic
gauge fields on optical lattices, which needless to say is an important
research topic.
In Ref. [\onlinecite{Jaksch2003}], Jaksch
and Zoller  proposed a theoretical scheme to generate artificial magnetic
fields on lattices.
Later on, this method has been generalized to generate non-Abelian $SU(N)$
gauge fields.
The single particle spectra exhibit a generalized Hofstadter butterfly
structure \cite{Osterloh2005}.
In addition, there has been a large experimental progess in the synthatic
effective magnetic fields in 2D lattices \cite{Aidelsburger2011,struck2012,
jimenez2012}.
More details about creating artificial gauge fields with neutral atoms and
some recent developments can be found in
Ref. [\onlinecite{Dalibard2011,tagliacozzo2012,goldman2013}].

\subsection{Tight-binding approximation and band parameters}
\label{eq:tight_bind}

\begin{figure}[htb]
\centering\epsfig{file=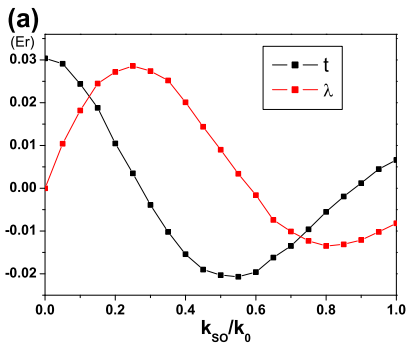,clip=1,width=0.7\linewidth}
\caption{ The dependence of the spin-independent hopping integral
$t$ and the spin-dependent one $\lambda$ {\it v.s.} the SO coupling
strength $k_{so}/k_0$. The optical potential depth is
$V_0=8E_r$.
From Ref. [\onlinecite{Cai2012}].
}
\label{fig:parameter}
\end{figure}

The tight-binding model in the square optical lattice will be derived
below for two-component bosons with synthetic SO coupling.
The single-particle Hamiltonian in the continuum is defined as
in Eq. (\ref{eq:single_rashba}) by replacing the trapping potential with
the periodic lattice potential as
\bea
V(x,y)=-V_0\Big[\cos^2 k_0x +\cos^2 k_0y \Big],
\label{eq:op}
\eea
where $k_0=2\pi/\lambda_0$ and the lattice constant $a=\lambda_0/2$.
The recoil energy is defined as $E_r=\hbar^2k_0^2/(2M)$.
For later convenience, the relative strength of SO coupling is quantified
by the dimensionless parameter $k_{so}/k_0$ with $k_{so}=M\lambda_R$.

A tight binding Hamiltonian for the lowest orbital band
with SO coupling can be written as \cite{Cai2012}
\bea
H&=&-\sum_{\langle ij\rangle,\sigma}
t_{ij; \sigma\sigma^\prime} \big[b^\dag_{i,\sigma} b_{j,\sigma'}+h.c \big]
+\sum_{i}[\frac{U}2 n_i^2-\mu n_i],
\nn \\
\label{eq:Hubbard}
\eea
where only the nearest neighbor hoppings are included.
$t_{ij;\sigma\sigma^\prime}$ can be decomposed into spin-independent
and spin-dependent components based on the following symmetry analysis
\bea
t_{ij;\sigma\sigma^\prime}= t+i\vec \lambda_{ij} \cdot \vec \sigma.
\eea
The coefficient of spin-dependent hopping is purely imaginary as a requirement
of the TR symmetry.
The optical lattice with the Rashba SO coupling possesses the reflection
symmetry with respect to the vertical plane passing bonds along the
$x$ and $y$-directions.
These reflection symmetries require that $\vec \lambda_{i,i+\hat e_x}
\parallel \hat e_y$, and $\vec \lambda_{i,i+\hat e_y}\parallel -\hat e_x$.
The four-fold rotation symmetry requires that $\vec \lambda_{i,i+\hat e_x}
\cdot \hat e_y = -\vec \lambda_{i,i+\hat e_y} \cdot \hat e_x$.
All these symmetry properties together constraint the spin-dependent hopping
up to a single parameter $\lambda$ as
\bea
\lambda_{i,i+\hat e_x}= \lambda \hat e_y, \ \ \,
\lambda_{i,i+\hat e_y}=-\lambda \hat e_x.
\eea

The band structure parameters $t$ and $\lambda$ are related to the overlap
integrals of the onsite SO coupled Wannier functions in neighboring sites.
In the case of a deep lattice, each site can be approximated by a local
harmonic potential.
The lowest energy Wannier states $\psi_{j_z=\pm\frac{1}{2}}$ are a
pair of Kramer doublets as presented in Eq. (\ref{eq:TR_doublet}).
The radial wavefunctions of the Wannier states $f(r)$ and $g(r)$ exhibit
the Friedel-type oscillations as explained in Sec. \ref{sect:2D_texture}.
Thus naturally $t$ and $\lambda$ should also exhibit such oscillations
as increasing the SO coupling parameter $k_{so}/k_0$.
This feature is numerically confirmed using the following method.
The tight-binding band spectra can be calculated easily as
\bea
E_{\pm}(\vec k)=\varepsilon(\vec k)\pm2\lambda\sqrt{\sin^2k_x+\sin^2k_y},
\label{eq:Energytb}
\eea
where $\varepsilon(\vec k)=-2t (\cos k_x + \cos k_y)$.
On the other hand, the band spectra can be calculated directly
from the continuum model with the lattice potential Eq.
(\ref{eq:op})
by using the basis of plane-waves.
By fitting these spectra using Eq. (\ref{eq:Energytb}), the values of $t$
and $\lambda$ are obtained and are plotted in Fig. \ref{fig:parameter}.
Both $t$ and $\lambda$ oscillate as increasing $k_{so}/k_0$, and the amplitudes
of their overall envelops decay.
For the spectra of Eq. (\ref{eq:Energytb}),
the square lattice breaks the rotational symmetry down into the 4-fold one,
thus the degeneracy of the Rashba ring is lifted.
The lower band has the 4-fold degenerate minima located at
$\vec Q=(\pm k, \pm k )$ with
\bea
k=\tan^{-1} \frac{1}{\sqrt 2}
\frac{\lambda}{t}.
\label{eq:spiral}
\eea

It should be pointed out that these lowest local Wannier states are
the eigenstates of the on-site total angular momentum with
$j_z=\pm\frac{1}{2}$.
Therefore, the eigen-bases defined by $(b_{i\uparrow}, b_{i\downarrow})^T$
should be those of $j_z$ not $\sigma_z$.
In the case of strong SO coupling $k_{so}\ge k_0$, the angular momenta
of these Wannier states nearly come from the orbital angular momentum moment,
while their spin moments are nearly average to zero.

If in the limit of very strong SO coupling such that $k_{so}\gg k_0$,
Landau level quantization effects appear within each site.
Many states with different values of $j_z$ are nearly degenerate as
presented in Sec. \ref{sect:landau}.
In this case, a single band model Eq. (\ref{eq:Hubbard})  fails even in
the case of the deep lattice.
It is justified only in
the case that $k_{so}/k\le 1$ in which the lowest Wannier states
are separated from others.

\subsection{Magnetic properties in the Mott-insulating state}
We consider the spin physics in the Mott-insulating phase of
Eq. (\ref{eq:Hubbard}).
In the simplest case, there is one particle per site with a two-fold
degenerate Kramer doublet.
The low energy superexchange Hamiltonian can be constructed
using the Schrieffer-Wolf transformation, which shows
the Dzyaloshinsky-Moriya (DM) type exchange
due to SO coupling \cite{dzyaloshinsky1958,moriya1960} as
\bea
H_{eff}=\sum_{i} H_{i, i+\hat{e}_x}+H_{i, i+\hat{e}_y},
\label{eq:eff1}
\eea
and
\bea
H_{i, i+\hat{e}_\mu}&=&-J_{1}
\vec S_{i}\cdot\vec S_{i+\hat{e}_\mu}-
J_{12}\vec d_{i,i+\hat{e}_\mu}\cdot
(\vec S_{i}\times\vec S_{i+\hat{e}_\mu})
\nn \\
&+&J_2[\vec S_{i}\cdot\vec {S}_{i+\hat{e}_\mu}
-2(\vec{S}_{i}\cdot\vec d_{i,i+\hat{e}_\mu})
(\vec{S}_{i+\hat{e}_\mu}\cdot \vec d_{i, i+\hat{e}_\mu})],\nn \\
\label{eq:DM1}
\eea
where $\hat e_\mu (\mu=x,y)$ are the unit vectors along
the $x$ and $y$-directions, respectively;
$J_1=4t^2/U$, $J_{12}=4t\lambda/U$, and $J_2=4\lambda^2/U$.
The DM vectors are defined as $\vec d_{i,i+\hat e_x}=\hat e_y$ and
$\vec d_{i,i+\hat e_y}=-\hat e_x$ which are perpendicular to each other.
This is similar to the case of the high T$_c$ cuprate
superconductors such as YBa$_2$Cu$_3$O$_6$ \cite{coffey1991,bonesteel1993}.
Consequently, these DM vectors in Eq. (\ref{eq:DM1}) cannot be removed by
gauge transformations, or, equivalently by varying local spin axes.
This brings frustrations to magnetic properties.
To obtain a qualitative understanding, two different
limits of $|\lambda|\ll |t|$ and $|\lambda|\gg |t|$ will be considered.

In the absence of SO coupling, i.e., $\lambda=0$, the system is in
the ferromagnetic state.
If $\lambda$ is small, the $J_2$-term brings the easy plane
anisotropy which prefers spin moments lie in the $xy$-plane. The
DM-vector further induces spin spiraling at a finite wavevector,
which can be shown by calculating the spin-wave spectra around the
variational ground states that spin moments lie along the
high symmetry line of diagonal directions, say, $[\bar{1}\bar{1}0]$.
The Holstein-Primakoff transformation is employed to transform
Eq. (\ref{eq:eff1}) into the magnon Hamiltonian,
\bea
H_{mg}=-J_0\sum_{i}\Big\{(\cos2\theta-i\frac{\sin2\theta}{\sqrt{2}})a_i^\dag
a_{i+e_x} \nn \\
+(\cos2\theta+i\frac{\sin2\theta}{\sqrt{2}})a_i^\dag
a_{i+e_y}+h.c \Big\},
\eea
where $a^\dagger$ is the creation operator for magnons deviating from the
$[110]$-direction; $\theta=\arctan(\lambda/t)$ as defined above.
We only keep quadric terms and ignore the terms proportional to $\sin^2\theta$
since $\lambda/t\ll 1$.
In momentum space, its spectra can be diagonalized as
\bea
\epsilon(\vec k)&=&-2 J_0  \Big\{ \cos2\theta
(\cos k_x +\cos k_y)\nn \\
&+&\frac{1}{\sqrt{2}}\sin2\theta
(\sin k_x-\sin k_y) \Big\},
\label{eq:spectrum2}
\eea
whose minima are located at $\vec Q_M=(2k, -2k)$ with
the value of $k$ given in Eq. (\ref{eq:spiral}).
This indicates that the ground state exhibits a
spin spiral order along the direction perpendicular to the quantized
axis in the spin-wave analysis.

Interestingly, in the opposite limit of $|\lambda/t|\gg 1$, Eq.
(\ref{eq:DM1}) can be related to that of $|\lambda/t|\ll 1$ through
a duality transformation.
On site $i$ with the coordinates $(i_x, i_y)$, $\vec S_i$ is transformed into
\bea
&&S_{i_x,i_y}^x\rightarrow
(-1)^{i_x}\mathbb{S}_{i_x,i_y}^x; \ \ \,
S_{i_x,i_y}^y\rightarrow
(-1)^{i_y}\mathbb{S}_{i_x,i_y}^y;\nn \\
&&S_{i_x,i_y}^z\rightarrow (-1)^{i_x+i_y}\mathbb{S}_{i_x,i_y}^z.
\label{eq:duality}
\eea
$\mathbb{\vec S}_i$ still maintains the spin commutation relation.
Under this transformation, the $J_1$-term transforms into
the $J_2$-term and vice versa, and the $J_{12}$-term is invariant.
Thus this dual transformation indicates that there is a one-to-one
correspondence between the $J_2$-dominant phase ($|\lambda/t|\gg 1$)
and that of $J_1$ with $|\lambda/t|\ll 1$ which has been analyzed above.

In the regime of intermediate values $\lambda/t$, a rich phase
diagram with different spin patterns appears.
Classical Monte Carlo simulations have been employed to calculate
the ground state phase diagram in current
literatures \cite{Cole2012,Radic2012,Gong2012}.
Various patterns have been found as a result of competition among
ferromagnetic exchange, easy-plane anisotropy, and the DM effect
induced spin spirals. These include the ferromagnetic,
antiferromagnetic, spiral, stripes, and vortex crystal orderings.
Furthermore, the superfluid-insulator transition for SO coupled
bosons has also been studied in Ref. [\onlinecite{Mandal2012}].
And similar topic in the presence of both SO coupling and effective
magnetic fields have also been considered in Ref. [\onlinecite{grass2011}].

\section{Conclusions}
\label{sect:conclusion}
We have reviewed unconventional BECs with SO coupling whose condensate
wavefunctions are complex-valued and are thus beyond the framework of
the ``no-node'' theorem.
Even at the single particle level, the spectra in harmonic traps exhibit
the structure of Landau-level like quantization induced by SO couplings.
Their energy dispersion is nearly flat with respect to angular momentum
in the case of strong SO coupling, and exhibit the Z$_2$-type topology.
The interacting condensates exhibit topologically non-trivial configurations.
In the 2D Rashba case, the spin density distributions are characterized by
the skyrmion type textures.
The 3D Weyl SO coupling induces the topological phase defects in the
quaternionic phase space, and the corresponding spin density
distributions are also non-trivial carrying non-zero
values of the Hopf-invariant.
In rotating traps, the condensate configurations are changed by vorticity
which results in a variety of structures including skyrmion lattices,
giant vortices, multi-domains of plane-waves.
In the strongly correlated Mott-insulating states, SO coupling exhibits
in the DM exchange interactions in the quantum magnetism.
The research of the novel states of SO coupled bosons is still in the
early stage.
In particular, the effect of SO couplings in the strong correlation
regime is still a largely unexplored field.
We expect that further exciting progress on the novel states of SO
coupled bosons will appear in the near future.

\acknowledgments

C. W. thanks I. Mondragon-Shem for early collaborations, and
L. Butov, T. L. Ho, H. Hu,  H. Pu, T.
Xiang, C. W. Zhang, F. Zhou, B. F. Zhu for helpful discussions.
X. F. Z.  acknowledges the support by NSFC (Grant Nos. 11004186),
National Basic Research Program of China 2011CB921204,
and the Strategic Priority Research Program of the Chinese Academy of Sciences
(Grant No. XDB01000000).
Y. L. and C. W. are supported by the NSF DMR-1105945 and AFOSR FA9550-11-1-
0067(YIP); Y.L. is also supported by the Inamori Fellowship.
Z.C.  acknowledges funding by DFG FOR 801.



\begin{thebibliography}{100}

\bibitem{ifmmodecheckZelsevZfiutiifmmodeacutecelsecfi2004}
I. \ifmmode \check{Z}\else \v{Z}\fi{}uti\ifmmode~\acute{c}\else \'{c}\fi{}, J.
  Fabian, and S. Das~Sarma, Rev. Mod. Phys. {\bf 76},  323  (2004).

\bibitem{Nagaosa2010}
N. Nagaosa {\it et~al.}, Rev. Mod. Phys. {\bf 82},  1539  (2010).

\bibitem{Xiao2010}
D. {Xiao}, M.-C. {Chang}, and Q. {Niu}, Rev. Mod. Phys. {\bf 82},
  1959  (2010).

\bibitem{Dyakonov1971}
M. D'yakonov and V. Perel, J. Exp. Theor. Phys. {\bf 13},  467  (1971).

\bibitem{Hirsch1999}
J.~E. Hirsch, Phys. Rev. Lett. {\bf 83},  1834  (1999).

\bibitem{Murakami2003}
S. {Murakami}, N. {Nagaosa}, and S.-C. {Zhang}, Science {\bf 301},  1348
  (2003).

\bibitem{Sinova2004}
J. {Sinova} {\it et~al.}, Phys. Rev. Lett. {\bf 92},  126603  (2004).

\bibitem{Hasan2010}
M.~Z. Hasan and C.~L. Kane, Rev. Mod. Phys. {\bf 82},  3045  (2010).

\bibitem{Qi2011}
X.-L. Qi and S.-C. Zhang, Rev. Mod. Phys. {\bf 83},  1057  (2011).


\bibitem{Galiski2013}
V. Galitski and I.B. Spielman, Nature \textbf{494}, 49-54 (2013).


\bibitem{Lin2009}
Y. Lin {\it et~al.}, Nature {\bf 462},  628  (2009).

\bibitem{Lin2009a}
Y. Lin {\it et~al.}, Phys. Rev. Lett. {\bf 102},  130401  (2009).

\bibitem{Lin2011}
Y. Lin, K. Jimenez-Garcia, and I. Spielman, Nature {\bf 471},  83  (2011).

\bibitem{Zhang2012a}
J. Zhang {\it et~al.}, Phys. Rev. Lett. {\bf 109},  115301  (2012).

\bibitem{Wang2012}
P. Wang {\it et~al.}, Phys. Rev. Lett. {\bf 109},  095301  (2012).

\bibitem{Qu2013}
C. {Qu} {\it et~al.}, arXiv:1301.0658  (2013).

\bibitem{cheuk2012}
L.~W. {Cheuk} {\it et~al.}, Phys. Rev. Lett. {\bf 109},  095302  (2012).

\bibitem{Feynman1972}
R.~P. Feynman, {\em Statistical Mechanics, A Set of Lectures} (Addison-Wesley
  Publishing Company, ADDRESS, 1972).

\bibitem{Wu2009}
C. Wu, Mod. Phys. Lett. B {\bf 23},  1  (2009).

\bibitem{Isacsson2005}
A. Isacsson and S.~M. Girvin, Phys. Rev. A {\bf 72},  053604  (2005).

\bibitem{Liu2006}
W.~V. Liu and C. Wu, Phys. Rev. A {\bf 74},  013607  (2006).

\bibitem{Kuklov2006}
A.~B. Kuklov, Phys. Rev. Lett. {\bf 97},  110405  (2006).

\bibitem{Cai2011}
Z. Cai and C. Wu, Phys. Rev. A {\bf 84},  033635  (2011).

\bibitem{Muller2007}
T. M\"uller, S. F\"olling, A. Widera, and I. Bloch, Phys. Rev. Lett. {\bf 99},
  200405  (2007).

\bibitem{Wirth2010}
G. Wirth, M. {\"O}lschl{\"a}ger, and A. Hemmerich, Nat. Phys. {\bf 7},  147
   (2010).

\bibitem{Olschlager2011}
M. \"Olschl\"ager, G. Wirth, and A. Hemmerich, Phys. Rev. Lett. {\bf 106},
  015302  (2011).


\bibitem{Wu2011}
C. Wu, I. Mondragon-Shem, arXiv:0809.3532V1;
C. Wu, I. Mondragon-Shem, and X.-F. Zhou, Chin. Phys. Lett. {\bf 28},
  097102  (2011).


\bibitem{Stanescu2008}
T.~D. Stanescu, B. Anderson, and V. Galitski, Phys. Rev. A {\bf 78},  023616
  (2008).

\bibitem{Jaksch2003}
D. Jaksch and P. Zoller, New J. Phys. \textbf{5}, 56 (2003).


\bibitem{Ruseckas2005}
J. Ruseckas, G. Juzeli\ifmmode~\bar{u}\else \={u}\fi{}nas, P. \"Ohberg, and M.
  Fleischhauer, Phys. Rev. Lett. {\bf 95},  010404  (2005).

\bibitem{Osterloh2005}
K. Osterloh {\it et~al.}, Phys. Rev. Lett. {\bf 95},  010403  (2005).

\bibitem{Campbell2011}
D.~L. Campbell, G. Juzeli\ifmmode~\bar{u}\else \={u}\fi{}nas, and I.~B.
  Spielman, Phys. Rev. A {\bf 84},  025602  (2011).

\bibitem{Juzeliunas2011}
G. Juzeliunas, J. Ruseckas, D. Campbell, and I. Spielman,  in {\em SPIE OPTO},
  International Society for Optics and Photonics (PUBLISHER, ADDRESS, 2011),
  pp.\ 79500M--79500M.

\bibitem{Dalibard2011}
J. Dalibard, F. Gerbier, G. Juzeli\ifmmode~\bar{u}\else \={u}\fi{}nas, and P.
  \"Ohberg, Rev. Mod. Phys. {\bf 83},  1523  (2011).

\bibitem{tagliacozzo2012}
L. Tagliacozzo, A. Celi, P. Orland, M. Lewenstein, arXiv:1211.2704
(2012).

\bibitem{goldman2013}
N. Goldman, F. Gerbier, and M. Lewenstein, arXiv:1301.4959v1 (2013).

\bibitem{Wang2010}
C. Wang, C. Gao, C. Jian, and H. Zhai, Phys. Rev. Lett. {\bf 105},  160403
  (2010).

\bibitem{Ho2011}
T. Ho and S. Zhang, Phys. Rev. Lett. {\bf 107},  150403  (2011).

\bibitem{Anderson2011}
B. Anderson, J. Taylor, and V. Galitski, Phys. Rev. A {\bf 83},  031602
  (2011).


\bibitem{Burrello2010}
M. Burrello and A. Trombettoni, Phys. Rev. Lett. \textbf{105}, 125304 (2010).

\bibitem{Burrello2011}
M. Burrello and A. Trombettoni, Phys. Rev. A \textbf{84}(4), 043625 (2011).


\bibitem{Kawakami2011}
T. Kawakami, T. Mizushima, and K. Machida, Phys. Rev. A {\bf 84},  011607
  (2011).

\bibitem{Li2011}
Y. Li and C. Wu, arXiv:1103.5422  (2011).

\bibitem{Sinha2011}
S. Sinha, R. Nath, and L. Santos, Phys. Rev. Lett. {\bf 107},  270401  (2011).

\bibitem{Xu2011a}
Z.~F. Xu, R. L\"u, and L. You, Phys. Rev. A {\bf 83},  053602  (2011).

\bibitem{Yip2011}
S.-K. Yip, Phys. Rev. A {\bf 83},  043616  (2011).

\bibitem{Zhu2011}
Q. Zhu, C. Zhang, and B. Wu, arXiv:1109.5811  (2011).

\bibitem{Anderson2012}
B. Anderson and C. Clark, arXiv:1206.0018  (2012).

\bibitem{Anderson2012a}
B.~M. Anderson, G. Juzeli\ifmmode~\bar{u}\else \={u}\fi{}nas, V.~M. Galitski,
  and I.~B. Spielman, Phys. Rev. Lett. {\bf 108},  235301  (2012).

\bibitem{Barnett2012}
R. Barnett {\it et~al.}, Phys. Rev. A {\bf 85},  023615  (2012).

\bibitem{Deng2012}
Y. Deng {\it et~al.}, Phys. Rev. Lett. {\bf 108},  125301  (2012).

\bibitem{Grass2012}
T. Gra\ss{}, B. Juli{\'a}-D{\'\i}az, and M. Lewenstein, arXiv:1210.8035  (2012).


\bibitem{He2012b}
P. He, R. Liao, and W. Liu, Phys. Rev. A {\bf 86},  043632  (2012).

\bibitem{Hu2012a}
H. Hu, B. Ramachandhran, H. Pu, and X. Liu, Phys. Rev. Lett. {\bf 108},  10402
  (2012).

\bibitem{LiYun2012}
Y. Li, L. Pitaevskii, and S. Stringari, Phys. Rev. Lett. {\bf 108},  225301
  (2012).

\bibitem{Ramachandhran2012}
B. Ramachandhran {\it et~al.}, Phys. Rev. A {\bf 85},  023606  (2012).

\bibitem{Ruokokoski2012}
E. Ruokokoski, J. Huhtam{\"a}ki, and M. M{\"o}tt{\"o}nen, arXiv:1205.4601
  (2012).

\bibitem{Sedrakyan2012}
T. Sedrakyan, A. Kamenev, and L. Glazman, arXiv:1208.6266  (2012).

\bibitem{Xu2012}
X. Xu and J. Han, Phys. Rev. Lett. {\bf 108},  185301  (2012).

\bibitem{Xu2012b}
Z. Xu, Y. Kawaguchi, L. You, and M. Ueda, Phys. Rev. A {\bf 86},  033628
  (2012).

\bibitem{Zhang2012}
D. Zhang, L. Fu, Z. Wang, and S. Zhu, Phys. Rev. A {\bf 85},  043609  (2012).

\bibitem{Zhang2012c}
X. Zhang {\it et~al.}, Phys. Rev. A {\bf 86},  063628  (2012).

\bibitem{Zhang2012d}
Y. Zhang, L. Mao, and C. Zhang, Phys. Rev. Lett. {\bf 108},  35302  (2012).

\bibitem{Zheng2012}
W. Zheng and Z. Li, Phys. Rev. A {\bf 85},  053607  (2012).

\bibitem{Cui2012}
X. {Cui} and Q. {Zhou}, arXiv:1206.5918  (2012).

\bibitem{Li2012c}
Y. Li, X. Zhou, and C. Wu, Phys. Rev. B {\bf 85},  125122  (2012).

\bibitem{Weyl1929}
H. Weyl, Z. Phys. {\bf 56},  330  (1929).

\bibitem{Li2012b}
Y. Li, X. Zhou, and C. Wu, arXiv:1205.2162  (2012).

\bibitem{Kawakami2012}
T. Kawakami, T. Mizushima, M. Nitta, and K. Machida, Phys. Rev. Lett. {\bf
  109},  015301  (2012).

\bibitem{Zhang2013}
D.-W. {Zhang} {\it et~al.}, arXiv:1301.2869  (2013).

\bibitem{Yao2008}
W. Yao and Q. Niu, Phys. Rev. Lett. {\bf 101},  106401  (2008).

\bibitem{High2011}
A. High {\it et~al.}, arXiv:1103.0321  (2011).

\bibitem{High2012}
A. High {\it et~al.}, Nature \textbf{483}, 584¨C588  (2012).

\bibitem{Ghosh2011}
S.~K. Ghosh, J.~P. Vyasanakere, and V.~B. Shenoy, Phys. Rev. A {\bf 84},
  053629  (2011).

\bibitem{Radic2011}
J. Radi{\'c}, T. Sedrakyan, I. Spielman, and V. Galitski, Phys. Rev. A {\bf
  84},  063604  (2011).

\bibitem{Xu2011}
X. Xu and J. Han, Phys. Rev. Lett. {\bf 107},  200401  (2011).

\bibitem{Liu2012}
C. Liu and W. Liu, Phys. Rev. A {\bf 86},  033602  (2012).

\bibitem{Zhao2013}
Y. Zhao, J. An, and C. Gong, Phys. Rev. A {\bf 87},  013605  (2013).

\bibitem{ZhouXF2011}
X. Zhou, J. Zhou, and C. Wu, Phys. Rev. A {\bf 84},  063624  (2011).


\bibitem{Ramachandhran2013}
B. Ramachandhran, H. Hu, and H. Pu, arXiv:1301.0800v1.

\bibitem{grass2012b}
T. Gra\ss{}, B. Juli\'{a}-D\'{\i}az, N. Barber\'{a}n, and M. Lewenstein, Phys. Rev. A \textbf{86}, 021603(R)
(2012).

\bibitem{grass2012c}
T. Gra\ss{}, B. Juli\'{a}-D\'{\i}az, M. Lewenstein, arXiv:1210.8035 (2012).

\bibitem{palmer2011}
R.~N. Palmer and J.~K. Pachos, New J. Phys. \textbf{13}, 065002 (2011).

\bibitem{komineas2012}
S. Komineas and N.~ R. Cooper, Phys. Rev. A \textbf{85}(5) 053623 (2012).

\bibitem{Cai2012}
Z. Cai, X. Zhou, and C. Wu, Phys. Rev. A {\bf 85},  061605  (2012).

\bibitem{Cole2012}
W. Cole, S. Zhang, A. Paramekanti, and N. Trivedi, Phys. Rev. Lett. {\bf 109},
  85302  (2012).

\bibitem{Radic2012}
J. Radi{\'c}, A. Di~Ciolo, K. Sun, and V. Galitski, Phys. Rev. Lett. {\bf 109},
   85303  (2012).

\bibitem{Gong2012}
M. Gong, Y. Qian, V. Scarola, and C. Zhang, arXiv:1205.6211
  (2012).

\bibitem{Mandal2012}
S. Mandal, K. Saha, and K. Sengupta, arXiv:1205.3178  (2012).


\bibitem{grass2011}
T. Gra\ss{}, K. Saha, K. Sengupta, and M.B. Lewenstein, Phys. Rev. A \textbf{84}, 053632 (2011).

\bibitem{Iskin2011}
M. Iskin and A. Suba{\c{s}}{\i}, Phys. Rev. Lett. {\bf 107},  50402  (2011).

\bibitem{Jiang2011}
L. Jiang, X. Liu, H. Hu, and H. Pu, Phys. Rev. A {\bf 84},  063618  (2011).

\bibitem{Zhou2011}
J. Zhou, W. Zhang, and W. Yi, Phys. Rev. A {\bf 84},  063603  (2011).

\bibitem{Doko2012}
E. Doko, A. Suba{\c{s}}{\i}, and M. Iskin, Phys. Rev. A {\bf 85},  053634
  (2012).

\bibitem{He2012}
L. He and X. Huang, Phys. Rev. Lett. {\bf 108},  145302  (2012).

\bibitem{He2012a}
L. He and X. Huang, arXiv:1207.2810  (2012).

\bibitem{Hu2012}
H. Hu {\it et~al.}, arXiv:1208.5841  (2012).

\bibitem{Liu2012a}
X. Liu, Phys. Rev. A {\bf 86},  033613  (2012).

\bibitem{Maldonado-Mundo2012}
D. Maldonado-Mundo, P. Ohberg, and M. Valiente, arXiv:1212.3565  (2012).

\bibitem{Martone2012}
G. Martone, Y. Li, L. Pitaevskii, and S. Stringari, Phys. Rev. A {\bf 86},
  063621  (2012).

\bibitem{Orth2012}
P. Orth {\it et~al.}, arXiv:1212.5607  (2012).

\bibitem{Vyasanakere2012}
J. Vyasanakere and V. Shenoy, arXiv:1201.5332  (2012).

\bibitem{Zhang2012b}
S. Zhang, X. Yu, J. Ye, and W. Liu, arXiv:1212.0424  (2012).

\bibitem{Li2012d}
Y. {Li}, S.-C. {Zhang}, and C. {Wu}, arXiv:1208.1562  (2012).

\bibitem{Li2012}
Y. Li, K. Intriligator, Y. Yu, and C. Wu, Phys. Rev. B {\bf 85},  085132
  (2012).


\bibitem{halperin1982}
B.~I. Halperin, Phys. Rev. B {\bf 25},  2185  (1982).

\bibitem{kane2005}
C.~L. Kane and E. J. Mele, Phys. Rev. Lett. \textbf{95}, 146802
(2005).

\bibitem{fu2007}
L. Fu and C.~L. Kane, Phys. Rev. B \textbf{76}, 045302 (2007).

\bibitem{moore2007}
J.~E. Moore and L. Balents, Phys. Rev. B \textbf{75}, 121306
(2007).

\bibitem{roy2010}
R. Roy, New J. Phys. \textbf{12}, 065009 (2010).



\bibitem{Zhou2003}
F. Zhou, Int. J. Mod. Phys. B {\bf 17},  2643  (2003).

\bibitem{adler1995}
S. Adler, {\em Quaternionic quantum mechanics and quantum fields} (Oxford
  University Press, USA, ADDRESS, 1995), Vol.~88.

\bibitem{finkelstein1962}
D. Finkelstein, J. Jauch, S. Schiminovich, and D. Speiser, J. Math. Phys. {\bf 3},  207  (1962).

\bibitem{balatsky1992}
A. Balatsky, arXiv preprint cond-mat/9205006  (1992).

\bibitem{wilczek1983}
F. Wilczek and A. Zee, Phys. Rev. Lett. {\bf 51},  2250  (1983).

\bibitem{nakahara2003}
M. Nakahara, {\em Geometry, topology and physics} (Taylor \& Francis, ADDRESS,
  2003).



\bibitem{kitaev2003}
A.~Y. Kitaev, Ann. Phys. \textbf{303}, 2-30 (2003).
\bibitem{Nayak2008}
C. Nayak, S.~H. Simon, A. Stern, M. Freedman, S. Das Sarma, Rev. Mod. Phys. \textbf{80}, 1083
(2008).



\bibitem{Aidelsburger2011}
M. Aidelsburger \emph{et al.}, Phys. Rev. Lett. \textbf{107}, 255301 (2011).

\bibitem{struck2012}
J. Struck  \emph{et al.}, Phys. Rev. Lett. \textbf{108}, 225304 (2012).

\bibitem{jimenez2012}
K. Jim\'{e}nez-Garc\'{\i}a \emph{et al.}, Phys. Rev. Lett. \textbf{108}, 225303 (2012).


\bibitem{dzyaloshinsky1958}
I. Dzyaloshinsky, J. Phys. Chem. SOLIDS {\bf 4},  241 (1958).

\bibitem{moriya1960}
T. Moriya, Phys. Rev. {\bf 120},  91  (1960).

\bibitem{coffey1991}
D. Coffey, T.~M. Rice, and F.~C. Zhang, Phys. Rev. B {\bf 44},  10112  (1991).

\bibitem{bonesteel1993}
N.~E. Bonesteel, Phys. Rev. B {\bf 47},  11302  (1993).




\end{thebibliography}

\end{document}